\documentclass[5p]{elsarticle}
\journal{Elsevier}

\usepackage{subfigure}
\usepackage{fancyhdr}
\usepackage{amsmath}
\usepackage{multirow}
\usepackage{amssymb }
\usepackage{color}
\usepackage{graphics} % for pdf,  bitmapped graphics files
\usepackage{graphicx}
\usepackage{amsmath}
\newtheorem{theorem}{\textbf{Theorem}}
\newtheorem{lemma}{\textbf{Lemma}}
\newtheorem{example}{\textbf{Example}}
\newtheorem{corollary}{\textbf{Corollary}}
\newtheorem{remark}{\textbf{Remark}}
\newtheorem{definition}{\textbf{Definition}}

\newtheorem{proposition}{\textbf{Proposition}}
\newtheorem{assumption}{\textbf{Assumption}}
\usepackage{url}

	\newenvironment{proof}{{{\bf Proof:}}}{\hfill $\square$\par}
	\usepackage{multirow} %multirow for format of table
	\usepackage{xcolor}
	\usepackage{multirow}
	\usepackage{setspace}
	\usepackage{url}
	\usepackage{subfigure}
	\usepackage{fancyhdr}
	\usepackage{amsmath}
	\usepackage{multirow}
	\usepackage{amssymb }
	\usepackage{color}
	\usepackage{graphics} % for pdf,  bitmapped graphics files
	\usepackage{graphicx}
	\usepackage{algorithm} %format of the algorithm
	\usepackage{algorithmic} %format of the algorithm
	\usepackage{subeqnarray}
	\usepackage{cases}
	\usepackage{threeparttable}

	 \normalsize
\begin{document}
		
		\begin{frontmatter}
			
			\title{On Polynomially Solvable Constrained Input Selections \\~ for Fixed and Switched Linear Structured Systems} %of Linear Systems
			%Controllability of Networked Relative Coupling Systems: a Structural Analysis

			%\author{Yuan Zhang and Tong Zhou$^{\dag}$% <-this % stops a space
				%\thanks{*This work was supported in part by the NNSFC under Grant 61573209 and 61733008. {\bf This work is just an extension of the conference paper \cite{Y_Zhang_2017}}.}% <-this % stops a space
				%\thanks{$^{\dag}$Yuan Zhang and Tong Zhou are with the Department of Automation and TNList, Tsinghua University, Beijing, 100084, P.~R.~China
					%        {(email: {\tt\small zhangyuan14@mails.tsinghua.edu.cn, tzhou@mail.tsinghua.edu.cn}).}}%
				%}
			
			\author{Yuan Zhang, Yuanqing Xia, Shenyu Liu, and Zhongqi Sun}%, Jinhui Zhang , and Yufeng Zhan %,yu-feng.zhan
			\address{School of Automation, Beijing Institute of Technology, Beijing, China\\~Email: $\emph{\{zhangyuan14,xia\_yuanqing, shenyuliu, zhongqisun\}@bit.edu.cn}$} %,zhangjinh
			%\thanks{This work was supported in part by the China Postdoctoral Innovative Talent Support Program under Grant BX20200055,  the China Postdoctoral
				%Science Foundation under Grant 2020M680016, and the National Natural Science Foundation of China under Grant 62003042. Corresponding author: Y. Xia}
			
			%\ead{{zhangyuan14@bit.edu.cn,xia_yuanqing@bit.edu.cn@bit.edu.cn}}
			\fntext[myfootnote]{This work was supported in part by the
				National Natural Science Foundation of China under Grant 62003042. } %zhangyuan14@bit.edu.cn, xia_yuanqing@bit.edu.cn@bit.edu.cn}

		\begin{abstract} This paper investigates two related optimal input selection problems for fixed (non-switched) and switched structured systems. More precisely, we consider selecting the minimum cost of inputs from a prior set of inputs, and selecting the inputs of the smallest possible cost with a bound on their cardinality, all to ensure system structural controllability. Those problems have attracted much attention recently; unfortunately, they are NP-hard in general. In this paper, it is found that, if the                                                                                                                                                                                                                                                                                                                                                                                                                                                                                                                                                                                                                                                                                                                                                                                                                                                                                                                                                                                                                                                                                                                                                                                                                                                                                                                                                                                                                                                                                                                                                                                                                                                                                                                                                                                                                                                                                                                                                                                                                                                                                                                                                                                                                                                                                                                                                                                                                                                                                                                                                                                                                                                                                                                                                                                                                                                                                                                                                                                                                                                                                                                                                                                                                                                                                                                                                                                                                                                                                                                                                                                                                                                                                                                                                                                                                                                                                                                                                                                                                                                                                                                                                                                                                                                                                                                                                                                                                                                                                                                                                                                                                                                                                                                                                                                                                                                                                                                                                                                                                                                                                                                                                                                                                                                                                                                                                                                                                                                                                                                                                                                                                                                                                                                                                                                                                                                                                                                                                                                                                                                                                                                                                                                                                                                                                                                                                                                                                                                                                                                                                                                                                                                                                                                                                                                                                                                                                                                                                                                                                                                     input structure satisfies certain `regularizations', which are characterized by the proposed restricted total unimodulairty notion, those problems can be solvable in polynomial time via linear programming (LP) relaxations. Particularly, the obtained characterizations depend only on the incidence matrix relating the inputs and the source strongly connected components (SCC) of the system structure,  irrespective of how the inputs actuate states within the same SCC. They cover all the currently known polynomially solvable cases (such as the dedicated input case), and contain many new cases unexploited in the past, among which the source-SCC separated input (SSSI) constraint is highlighted. Further, for switched systems, the obtained polynomially solvable condition (namely the joint SSSI constraint) does not require each of the subsystems to satisfy the SSSI constraint. We achieve these by first formulating those problems as equivalent integer linear programmings (ILPs), and then proving {\emph{total unimodularity}} of the corresponding constraint matrices. This property allows us to solve those ILPs efficiently via LP-relaxation. We also discuss solutions obtained via LP-relaxation and LP-rounding in the general case. Several examples are given to illustrate the obtained theoretical results.

		\end{abstract}
		
		\begin{keyword}
			Structural controllability, input selection, switched system, linear programming, total unimodularity
		\end{keyword}
		
	\end{frontmatter}
	%\keywords{Controllability robustness, actuator removals, security analysis, recursive tree, submodular function minimization, complexity}

	%%%%%%%%%%%%%%%%%%%%%%%%%%%%%%%%%%%%%%%%%%%%%%%%%%%%%%%%%%%%%%%%%%%%%%%%%%%%%%%%
	\section{Introduction} \label{intro-sec}		
Over the past decade, input/output (I/O) selections for a large-scale dynamic system to achieve certain performances have received considerable attention \cite{Y.Y.2011Controllability,A.Ol2014Minimal,T2016On,P.Fa2014Controllability}. Examples include estimation error minimization of the Kalman filter by sensor placement \cite{zhang2017sensor}, stabilization by joint I/O selection and feedback design \cite{nugroho2019algorithms}, achieving various performances related to controllability/observability \cite{A.Ol2014Minimal,T2016On,P.Fa2014Controllability,zhang2019structural}, etc. This paper is about I/O selections for controllability. % convergence error minimization of multi-agent systems \cite{clark2014minimizing} by leader selection,

Broadly speaking, problems concerning I/O selections for controllability can be divided into two categories. The first one is
selecting inputs to optimize some control energy-related metrics, such as the trace, determinant, or inverse of the minimum eigenvalue of the controllability Gramian \cite{T2016On,P.Fa2014Controllability,ikeda2018sparsity}. One typical approach to some of those problems is exploiting the modular, submodular, or weak submodular structure of the corresponding optimization problems, which often leads to greedy algorithms with provable approximation guarantees \cite{T2016On}. The second one is to design certain `sparse' inputs for ensuring controllability in the qualitative sense. Depending on what qualitative notion is adopted, this category can also be divided into two subclasses. When the purpose is to ensure controllability in the numerical sense, the minimum number of inputs needed has an analytical expression (i.e., being the maximum geometric multiplicity of the system state transition matrix) \cite{Y_Zhang_2018}. However, if the available input vectors are given a priori, this problem turns out to be NP-hard \cite{A.Ol2014Minimal}. The other subclass is about structural controllability, an alternative notion of controllability in the generic sense \cite{C.Co2013Input}, detailed as follows.

%if each input vector can be constructed arbitrarily

The problems of optimally selecting inputs to achieve structural controllability can be roughly classified into two classes, depending on the objectives. The first class of problems aims at determining the minimum number/cost of input links (typically, the sparsest input matrices) for structural controllability.  When there is no constraint on the structure of the input configuration or all the available inputs are dedicated (i.e., each input can actuate at most one state variable), it has been shown that these problems can be solved in polynomial time by transforming them to some maximum matching problems \cite{A_Olshevsky_2015,pequito2016minimum}. Recently, \cite{ACC2022-zhang} extends the dedicated input constraint to the so-called source strongly-connected component (SCC) grouped constraint and shows that if the available inputs satisfy this constraint, then the aforementioned problems are polynomially solvable. Further, \cite{zhang2019minimal} shows that finding the sparsest interconnection structure (both among states and between inputs and states) for a system to be structurally controllable is NP-hard, if the available interconnections among states are given a priori. 

The second class of problems intends to find the minimum number/cost of inputs (typically, the input matrices with the smallest number of columns) to achieve structural controllability. Note that compared to the first class of problems, selecting an input indicates all the input links incident to this input are selected simultaneously. It has been shown this problem has an analytical solution if there is no
constraint on the input structure \cite{Y.Y.2011Controllability}.  However, if the available inputs are given a priori (called the {\emph{constrained input selection problem}} in this case), this problem is generally NP-hard \cite{pequito2015complexity}. Due to the NP-hardness, only very limited scenarios are known for this problem to be polynomially solvable, such as the dedicated input case in \cite{A_Olshevsky_2015,pequito2016minimum}. For approximation, \cite{pequito2015complexity} reduces this problem to the set-cover problem for a special case, and \cite{moothedath2018flow} gives a flow-network based approximation algorithm.% where the approximation factor is determined by the maximum degree of input vertices in the constructed flow network.

In this paper, we advance the state of the art in searching polynomially solvable conditions for constrained input selection problems. 
More precisely, in addition to considering the problem of selecting the minimum cost of inputs from a given set to achieve system structural controllability, we also reinvestigate this problem by imposing an upper bound on the number of selected inputs. To our knowledge, no polynomial-time algorithms have been reported for these problems in the non-dedicated input case, except for some trivial cases (c.f., the system structure is strongly-connected). The initial idea of our study is that, since the addressed problems are NP-hard due to the fact that determining the minimum number of inputs to achieve input-reachability is NP-hard \cite{pequito2015complexity} (see also Section \ref{preliminary}) , what happens for the class of systems where the latter problem is no longer intractable?

%(the obtained problem is called the cardinality-constrained minimum cost input selection problem)
%Although this constraint covers almost all the past known polynomially solvable cases, it is still not enough to answer that open problem. In this paper, we re-investigate the constrained input selection problems. Instead of giving approximation algorithms, we explore conditions under which those problems are polynomially solvable. , and selecting the inputs with the smallest possible cost but with a bound on their cardinality, all to ensure system structural controllability.  

{\bf Main contributions:} Starting from this point, we reveal if the input structure satisfies the so-called source-SCC separated input constraint (SSSI constraint), i.e., no inputs can actuate two different source-SCCs simultaneously, the addressed input selection problems can be solved in polynomial time. We further generalize this result, showing that if the input structure satisfies certain `regularizations', which are characterized by the restricted total unimodularity (TU) notion introduced in this paper, those problems are polynomially solvable. The obtained conditions largely extend the currently known polynomially solvable ones, as they depend only on the connections between the inputs and the source-SCCs of the system structure, but irrespective of how each input actuates states within the same SCC.  Those results are further extended to the switched systems for similar input selection problems. Particularly, a joint SSSI constraint is proposed, which does not require each subsystem to meet the SSSI constraint but preserves the polynomial solvability. Key to our results is first formulating the constrained input selection problems as equivalent integer linear programming (ILP) problems, and then proving that the corresponding constraint matrices are TU under the addressed conditions. This allows us to solve those ILPs efficiently by simply solving their linear programming (LP) relaxations. In this way, we provide an LP-based method for these problems with polynomial time complexity, conceptually different from the graph-theoretic ones.  For the general case, we study solutions obtained via LP-relaxation and LP-rounding, resulting in some lower and upper bounds for the minimum cost input selection problem. In particular, the lower bound is tighter than the one obtained via the graph-theoretic method, while the upper bound has a provable approximation factor for a special case. 	{We remark that the LP-based method has also been used in \cite{ACC2022-zhang} for solving a different input selection problem from this paper. Relative to \cite{ACC2022-zhang}, the considered problems here are essentially NP-hard and more intricate as the set cover problem is embedded. The obtained polynomial solvability conditions are also wider than that in \cite{ACC2022-zhang}.}  Partial results of this paper are scheduled to appear in \cite{CDC2022-zy}. While \cite{CDC2022-zy} only covers the SSSI constraint without proofs, this paper generalizes it to the restricted TU condition, provides the full proofs, and presents extensions to the switched systems as well as solutions in the general case.

%	input-source-SCC incidence matrix, namely, the restricted total unimodularity (TU). This characterizes  

%	This condition defines a large class of systems in which the source-SSCs may have certain autonomy/independence so that they do not receive control signals from the same input. 

The rest of this paper is organized as follows. Section \ref{problem-formulation} gives the problem formulations, and Section \ref{preliminary} provides some preliminaries in graph theory and structured systems. Section \ref{main-ILP} gives the ILP formulations of the considered problems, while Section \ref{main-result} presents the polynomially solvable conditions for the fixed (non-switched) systems. Solutions in the general case via LP-relaxation and LP-rounding are discussed in Section \ref{general-case}. Extensions to the switched systems are given in Section \ref{switched-case}.  Section \ref{example-sec} provides two illustrative examples. The last section concludes this paper.

Throughout this paper, for two vectors $a$ and $b$, $a\le b$ means $a_i\le b_i$ entry-wisely. A vector $a$ is integral if every element is an integer. For an optimization problem $\min \{\varphi(x): x\in \Lambda\}$, $\Lambda$ is called the feasible region, $x\in \Lambda$ is called a feasible solution, the minimum of the objective $\varphi(x)$ on $x\in \Lambda$ is called the optimal (objective) value, or optimum, while the $x$ for which the optimum is attained is an optimal solution. $1_{n\times m}$ ($0_{n\times m}$) denotes the $n\times m$ matrix with all entries $1$ ($0$). The set of positive integers is denoted as ${\mathbb N}_{+}$.

%	{\bf Notations and terminologies:} For two vectors $a$ and $b$, $a\le b$ means $a_i\le b_i$ entry-wisely. We say a vector $a$ is integral, if its every element is an integer. For an optimization problem $\min \{\varphi(x): x\in \Lambda\}$, $\Lambda$ is the feasible region, $x\in \Lambda$ is a feasible solution, the minimum of the objective $\varphi(x)$ on $x\in \Lambda$ is called the optimal (objective) value, or optimum, while the $x$ for which the optimum is attained is called an optimal solution. $1_{n\times m}$ ($0_{n\times m}$) denotes the $n\times m$ matrix with all entries $1$ ($0$).

{		
	\section{Problem formulations} \label{problem-formulation}
	Consider a linear-time invariant system as
\begin{equation} \label{plant}
	\dot x(t)=\tilde A x(t) + \tilde B u(t),
\end{equation}
in which $x(t)\in {\mathbb R}^n$, $u(t)\in {\mathbb R}^m$ are the state variables and inputs, and $\tilde A\in {\mathbb R}^{n\times n}$, $\tilde B\in {\mathbb R}^{n\times m}$.

A structured matrix is a matrix with entries being either fixed zero or a free parameter. Denote the set of $n_1\times n_2$ structured matrices by $\{0,*\}^{n_1\times n_2}$, where $*$ represents the free parameters. Let $A$ and $B$ be two structured matrices specifying the sparsity patterns of $\tilde A$ and $\tilde B$, i.e., $ A_{ij}= 0$ implies $\tilde A_{ij}= 0$, and $ B_{ij}= 0$ implies $\tilde B_{ij}= 0$. In this way, $(\tilde A,\tilde B)$ is called a realization of $(A,B)$. 

\begin{definition} \cite{generic}
	$(A,B)$ is said to be structurally controllable, if there is a realization of it that is controllable.
\end{definition}

Controllability of system (\ref{plant}) is a generic property in the sense that, if $(A,B)$ is structurally controllable,  then almost all of its realizations are controllable; otherwise, none is controllable \cite{Lin_1974}. 		Given $B \in \{0,*\}^{n\times m}$ and ${\cal J}\subseteq \{1,...,m\}$,  let $B({\cal J})$ be the sub-matrix of $B$ consisting of columns indexed by ${\cal J}$.
Assign a non-negative {\emph{rational}} cost $c_{i}$ to each column of $B$, representing the cost of activating the $i$th input. We say $B$ is {\emph{dedicated}}, if each column of $B$ has at most one nonzero entry. With the notations above, we first consider the following optimal input selection problem:

%				Let $A$ and $B$ be structured matrices that characterize the sparsity patterns of $\tilde A$ and $\tilde B$, that is, $A_{ij}=0$ (resp. $B_{ij}=0$) implies $\tilde A_{ij}=0$ (resp. $\tilde B_{ij}=0$), for all $1\le i,j \le n$ (resp. $1\le i \le n, 1\le j \le m$). We may use $\{0,*\}^{n_1\times n_2}$ to denote the set of all structured matrices with the dimension $n_1\times n_2$, in which $0$ denotes the fixed zero entries, and $*$ the entries that can take values freely. For a structured matrix $M\in \{0,*\}^{n_1\times n_2}$, ${\cal S}(M)$ denotes the set of its realizations, i.e., ${\cal S}(M)=\{\tilde M\in {\mathbb R}^{n_1\times n_2}: \tilde M_{ij}=0 \ {\rm if} \ M_{ij}=0\}$. $(A,B)$ is said to be structurally controllable, if there exists $\tilde A\in {\cal S}(A)$ and  $\tilde B \in {\cal S}(B)$, so that $(\tilde A, \tilde B)$ is controllable. It is well-known that controllability is a generic property, in the sense that if $(A,B)$ is structurally controllable, then almost all of its realizations are controllable.

Problem ${\cal P}_1$: minimum cost input selection
\begin{equation}\begin{array}{l}
		\min \limits_{{\cal J}\subseteq \{1,...,m\}} \sum_{i\in {\cal J}} c_i \tag{${\cal P}_1$}\label{A1} \\
		{\rm s.t.} \ (A,B({\cal J})) \ {\rm structurally \ controllable}
\end{array}\end{equation}

That is, ${\cal P}_1$ seeks to select the inputs from the prior input matrix $B$ with the minimum total cost to achieve structural controllability. Next, we consider the following problem by adding a cardinality upper bound $k\in {\mathbb N}_+$ to the number of inputs:

Problem ${\cal P}_2$: cardinality-constrained minimum cost input selection
\begin{equation}\begin{array}{l}
		\min \limits_{{\cal J}\subseteq \{1,...,m\}} \sum_{i\in {\cal J}} c_i \tag{${\cal P}_2$}\label{A3} \\
		{\rm s.t.} \ (A,B({\cal J}))\ {\rm structurally \ controllable} \\
		\ \ \ \ \ |{\cal J}|\le k
\end{array}\end{equation}

In other words, ${\cal P}_2$ intends to select the inputs with a bound on their cardinality and with the total cost as small as possible to ensure structural controllability. It may happen that the optimal solution to ${\cal P}_1$, denoted by ${\cal J}^*$, has a much larger cardinality $|{\cal J}^*|$ than that to ${\cal P}_2$. Therefore, ${\cal P}_2$ may be desirable, for example, when the activation of new inputs may be more expensive compared to increasing the input costs.  Throughout this paper, without losing any generality, the following assumption is adopted, which is necessary for the feasibility of ${\cal P}_1$ and ${\cal P}_2$. 

\begin{assumption} \label{sc-assump}
	$(A,B)$ is structurally controllable.
\end{assumption}

\begin{remark}
	A related problem to ${\cal P}_2$ is selecting the set of inputs to achieve structural controllability with the smallest possible cost, meanwhile the cardinality is no more than any number of inputs ensuring system structural controllability, i.e., $k$ in ${\cal P}_2$ equals the optimum of ${\cal P}_1$ with unit input cost (i.e, $c_i=1, \forall i$; denote this value by $N^*_{{\cal P}_1}$). This problem is equivalent to the following one
	\begin{equation}\begin{array}{l}
			\min \limits_{{\cal J}\subseteq \{1,...,m\}} \sum_{i\in {\cal J}} (c_i+\gamma)  \tag{${\cal P}_{3}$} \\
			{\rm s.t.} \ (A,B({\cal J})) \ {\rm structurally \ controllable}
	\end{array} \end{equation}
	where $\gamma\doteq mc_{\max}$, with $c_{\max} \doteq \max_{1\le i \le m} c_i>0$. Here, $\gamma$ is the regularization parameter to penalize the cardinality of the solution,
	such that for any feasible solution with the cardinality larger than $N^*_{{\cal P}_1}$, its decrease in the cost (less than $mc_{\max}$) will not exceed the
	increase (at least $\gamma$) caused by the cardinality penalty. Hence, ${\cal P}_{3}$ is indeed a special case of ${\cal P}_{1}$.
\end{remark}

Next, consider the following switched linear system
\begin{equation} \label{plant-switched}
	\dot x(t)=\tilde A_{\sigma(t)} x(t) + \tilde B_{\sigma(t)} u(t),
\end{equation}
where $x(t)\in {\mathbb R}^n$ is the state, $u(t)\in {\mathbb R}^{m}$ is the piecewise continuous input, $\sigma(t): [0,\infty) \to \{1,...,p\}$ is the switching signal, $p$ is the number of the 
switching modes, and $(\tilde A_i, \tilde B_i)$ is called a subsystem (mode) of system (\ref{plant-switched}). $\sigma(t)=i$ implies the subsystem $(\tilde A_i, \tilde B_i)$ is activated as the system realization at time instant $t$, $i=1,2,...,p$.  We denote system (\ref{plant-switched}) as the pair
$(\tilde A_{\sigma(\cdot)}, \tilde B_{\sigma(\cdot)})$. {\emph{System (\ref{plant-switched}) is said to be controllable, if for any two states $x_0, x_f\in {\mathbb R}^n$, there exists a finite $t_f$, a switching signal $\sigma(t): [0, t_f)\to \{1,...,p\}$ and an input $u(t): [0, t_f)\to {\mathbb R}^m$, such that $x(0)=x_0$ and $x(t_f)=x_f$ \cite{Z.S2002Controllability}.}}

Similar to the above, let $A_{i}$ and $B_{i}$ be structured matrices specifying the sparsity patterns of $\tilde A_i$ and $\tilde B_i$, $i=1,...,p$, and we obtain a structured switched system corresponding to (\ref{plant-switched}), denoted by $(A_{\sigma(\cdot)}, B_{\sigma(\cdot)})$.  In this way, $(\tilde A_{\sigma(\cdot)}, \tilde B_{\sigma(\cdot)})$ is called a realization of $(A_{\sigma(\cdot)}, B_{\sigma(\cdot)})$. We call the nonzero columns of $B_i$ the effective input vectors, and suppose there are $m_i$ effective input vectors in $B_i$, $i=1,...,p$. Without loss of generality, assume those effective input vectors locate at the first $m_i$ columns of $B_i$. Then, we also use $(A_1,\cdots, A_p, B_1',\cdots, B_p')$ to denote system $(A_{\sigma(\cdot)}, B_{\sigma(\cdot)})$, where $B_i'$ consists of the effective input vectors of $B_i$.

\begin{definition} \cite{LiuStructural}
	The pair $(A_{\sigma(\cdot)}, B_{\sigma(\cdot)})$ is said to be structurally controllable, if there exists a realization that is controllable. 
\end{definition} 

Controllability of the switched system (\ref{plant-switched}) is again a generic property, characterized by structural controllability of $(A_{\sigma(\cdot)}, B_{\sigma(\cdot)})$. Similar to the non-switched case, assign a non-negative rational cost $c_{ij}$ to the $j$th column of $B_i$, representing the cost of using the $j$th input of the $i$th mode, $i=1,...,p$, $j=1,...,m_i$. 
Corresponding to ${\cal P}_1$ and ${\cal P}_2$, we consider the following two input selection problems for the switched system (\ref{plant-switched}) to achieve structural controllability:

Problem ${\cal P}_4$: minimum cost switched input selection 
\begin{equation}\begin{array}{l}
		\min \limits_{{\cal J}_i\subseteq \{1,...,m_i\}, i=1,...,p} \sum_{i=1}^p \sum_{j\in {\cal J}_i} c_{ij} \tag{${\cal P}_4$}\label{A1} \\
		{\rm s.t.} \ (A_1,\cdots, A_p, B_1({\cal J}_1),\cdots, B_p({\cal J}_p)) \\ \ {\rm structurally \ controllable}
\end{array}\end{equation}

Problem ${\cal P}_5$: cardinality-constrained minimum cost \\~ switched input selection ($k$ given)
\begin{equation}\begin{array}{l}
		\min \limits_{{\cal J}_i\subseteq \{1,...,m_i\}, i=1,...,p} \sum_{i=1}^p \sum_{j\in {\cal J}_i} c_{ij} \tag{${\cal P}_5$}\label{A1} \\
		{\rm s.t.} \ (A_1,\cdots, A_p, B_1({\cal J}_1),\cdots, B_p({\cal J}_p)) \\ {\rm structurally \ controllable} \\
		\ \ \ \ \ \sum \nolimits_{i=1}^p|{\cal J}_i|\le k. 
\end{array}\end{equation}

It is known that ${\cal P}_1$ is NP-hard in general \cite{pequito2015complexity}. Hence, ${\cal P}_2$ is NP-hard (by setting $k=m$, ${\cal P}_2$ reduces to ${\cal P}_1$), and so are with ${\cal P}_4$ and ${\cal P}_5$ (by setting $p=1$, ${\cal P}_4$ and ${\cal P}_5$ reduce to ${\cal P}_1$ and ${\cal P}_2$, respectively). Problems ${\cal P}_4$ and ${\cal P}_5$ differ from the input selection problem in \cite{pequito2017structural}, which imposes no cost, no prior constraint, and no cardinality upper bound on the available inputs.  To our knowledge, except for some special cases (such as the dedicated input case), no polynomially solvable conditions for those problems have been reported. 

Though the NP-hardness, in this paper we reveal that if the input structure satisfies certain `regularizations', all the above-mentioned problems are polynomially solvable. Particularly, the obtained results are state-of-the-art, in the sense that they cover all the currently known polynomially solvable cases for those problems, including the dedicated input case, and contain many new cases unexploited in the past. Our main tool is the LP-relaxation, which also provides an alternative method for those problems apart from the traditional graph-theoretic one. We will also discuss the solutions obtained by LP-relaxation and LP-rounding in the general case.

	%	   The purpose of this paper is to characterize a large class of systems associated with which ${\cal P}_1$ and ${\cal P}_2$ are polynomially solvable. This characterization contains the dedicated input case as a special one. Our tool is the LP-relaxation, i.e., formulating those problems as ILPs and showing their corresponding LP-relaxations have integral optimal solutions for the specific class of systems characterized in this paper. We will also discuss the solutions obtained by LP-rounding in the general case.   %{\color{red} For ${\cal P}_1$ with uniform cost $\{c_i\}$, we will additionally provide the corresponding graph-theoretic algorithm. }

	\section{Preliminaries} \label{preliminary}
Some preliminaries in graph theory and structured systems are introduced for the subsequent derivations. These results are quite standard, and readers can refer to \cite{generic}.

A directed graph (digraph) is denoted by $G=(V,E)$, in which $V$ is the vertex set and $E\subseteq V\times V$ is the edge set. A path in a digraph is a sequence of edges, in which the terminal vertex of the preceding edge is the starting vertex of the successive edge. If there is a path from vertex $v_j$ to vertex $v_i$, we say $v_i$ is {\emph{reachable}} from $v_j$. A digraph is said to be strongly connected if any pair of its vertices are reachable from each other. An SCC of a digraph is its subgraph that is strongly connected, and no edges or vertices can be included in this subgraph without breaking the property of being strongly connected. We say a vertex {\emph{connects with}} a subgraph, if there is an edge from this vertex to a vertex of this subgraph. A bipartite graph, denoted by $G=(V_L,V_R,E_{RL})$, is a graph whose vertices can be partitioned into two disjoint parts $V_L$ and $V_R$, such that no edges of $E_{RL}$ have two end vertices within the same part.  A matching of a bipartite graph is a set of edges, among which any two do not share a common end vertex. A vertex is matched w.r.t. a matching, if it is an end vertex of an edge in this matching. The maximum matching is the matching with as many edges as possible. A {\emph{perfect matching}} of $G$ is a matching that matches every vertex of $G$ (implying $|V_L|=|V_R|$). 

For $A\in \{0,*\}^{n\times n}$, $B\in \{0,*\}^{n\times m}$, the state digraph is ${\cal G}(A)=(X,E_{A})$, with $X=\{x_1,...,x_n\}$ the set of state vertices, and $E_{A}=\{(x_j,x_i): A_{ij}\ne 0\}$ the set of state edges. The system digraph is ${\cal G}(A,B)=(X\cup U, E_{A}\cup E_{B})$ with the input vertices $U=\{u_1,...,u_m\}$ and the input links (edges) $E_{B}=\{(u_i,x_j): B_{ji}\ne 0\}$. Moreover, the bipartite graph associated with $(A,B)$ is defined as ${\cal B}(A,B)=(X_L,U\cup X_R, E_{XX}\cup E_{UX})$, in which $X_L=\{x^L_1,...,x^L_n\}$, $X_R=\{x_1^R,...,x_n^R\}$ are copies of $X$, $U=\{u_1,...,u_m\}$, $E_{XX}=\{(x^R_j,x^L_i): A_{ij}\ne 0\}$, and $E_{UX}=\{(u_j,x^L_i): B_{ij}\ne 0\}$. Define ${\cal B}(A)$ as ${\cal B}(A)\doteq (X_L,X_R,E_{XX})$.

Decompose ${\cal G}(A)$ into SCCs,  and suppose the $i$th SCC has a vertex set $X_i\subseteq X$ ($1\le i \le n_c$, with $n_c$ being the number of SCCs). An SCC is called a {\emph{source-SCC}}, if in ${\cal G}(A)$, there is no incoming edge to vertices of this SCC from other SCCs; otherwise, it is called a {\emph{non-source-SCC}}.   Suppose there are $r$ source-SCCs in ${\cal G}(A)$, with the set of their indices being ${\cal I}\doteq \{1,...,r\}$, $1\le r \le n_c$. A state vertex $x_i\in X$ is said to be {\emph{input-reachable}}, if it is reachable from an input vertex $u\in U$ in ${\cal G}(A,B)$. With those notions, the following lemma characterizes structural controllability.

%	For each $i\in {\cal I}$, let $X_i^L=\{x^L_j\in X_L: x_j\in X_i\}$, and $E_i=\{(u,x)\in E_{UX}: x\in X_i^L, u\in U\}$ as the set of input links between $U$ and $X^L_i$ in ${\cal B}(A,B)$. A state vertex $x_i\in X$ is said to be {\emph{input-reachable}}, if it is reachable from an input vertex $u\in U$ in ${\cal G}(A,B)$. %With some abuse of terminology, if every vertex in $X_i$ is input-reachable in ${\cal G}(A,B)$, we say $X_i^L$ is input-reachable in ${\cal B}(A,B)$. With those notions, the following lemma characterizes structural controllability.

\begin{lemma}[\cite{generic}] \label{sc-theory} $(A,B)$ is structurally controllable, if and only if the following two conditions hold simultaneously:
	
	i) every state vertex $x_i\in X$ is input-reachable;
	
	ii) there is a maximum matching in ${\cal B}(A,B)$ so that every $x_i^L\in X_L$ is matched.
\end{lemma}

%	By the definition of input-reachability, it is obvious condition i) of {Lemma \ref{sc-theory}} is equivalent to that, $E_i\ne \emptyset$ for each $i\in {\cal I}$.

	\section{ILP formulations of ${\cal P}_1$ and ${\cal P}_2$} \label{main-ILP}
	
	In this section, we formulate problems ${\cal P}_1$ and ${\cal P}_2$ as equivalent ILPs.
	
	%		In our ILP formulations, we introduce two binary variables $y=\{y_{uv}: (u,v)\in E_{XX}\cup E_{UX}\}$ and $t=\{t_i: i\in {\cal U}\}$, where ${\cal U}\doteq \{1,...,m\}$. In a feasible solution $(y,t)$ to ${\cal P}_i$ ($i=1,2$), $y_{uv}=1$ indicates the edge $(u,v)\in E_{XX}\cup E_{UX}$ is in a particular maximum matching of ${\cal B}(A,B)$, and $y_{uv}=0$ means the contrary. For $i\in {\cal U}$, $t_i=1$ means input $u_i$ is selected, while $t_i=0$ the contrary. To present the ILP formulations, matrix $w=[w_{ij}]\in \{0,1\}^{r\times m}$ is introduced as follows:  $w_{ij}=1$ if $(u_j,x_l)\in E_{UX}$ for some $x_l\in X_i$, and $w_{ij}=0$ if no such $x_l$ exists. In other words, $w_{ij}=1$ if and only if input $u_j$ directly connects with the source-SCC $X_i$. Let $E_{u_j}=\{(u_j,v): (u_j,v)\in E_{UX}\}$ be the set of input links incident to $u_j$, $\forall j\in {\cal U}$.

	In our ILP formulations, we introduce two binary variables $y=\{y_{uv}: (u,v)\in E_{XX}\cup E_{UX}\}$ and $t=\{t_i: i\in {\cal U}\}$, where ${\cal U}\doteq \{1,...,m\}$. In a feasible solution $(y,t)$ to the corresponding ILPs, $y_{uv}=1$ indicates the edge $(u,v)\in E_{XX}\cup E_{UX}$ is in a particular maximum matching of ${\cal B}(A,B)$, and $y_{uv}=0$ means the contrary. For $i\in {\cal U}$, $t_i=1$ means input $u_i$ is selected for the corresponding ${\cal P}_j$ ($j=1,2$), while $t_i=0$ the contrary.	To present the ILP formulations, the {\emph{source-SCC-input incidence matrix}} $w=[w_{ij}]\in \{0,1\}^{r\times m}$ is introduced as follows:  $w_{ij}=1$ if $(u_j,x_l)\in E_{UX}$ for some $x_l\in X_i$, and $w_{ij}=0$ if no such $x_l$ exists. In other words, $w_{ij}=1$ if and only if input $u_j$ directly connects with the source-SCC $X_i$. Let $E_{u_j}=\{(u_j,v): (u_j,v)\in E_{UX}\}$ be the set of input links incident to $u_j$, $j\in {\cal U}$.

	\begin{proposition} \label{ILP-formulation} Under Assumption \ref{sc-assump}, ${\cal P}_1$ is equivalent to the following ILP ${\cal P}^{\rm ILP}_1$ in the sense that, for an optimal solution $(y^\star, t^\star)$ to ${\cal P}_1^{\rm ILP}$, $S^\star=\{u_i: t^\star_i=1, i\in {\cal U}\}$ is an optimal solution to ${\cal P}_1$.
		\begin{align}
			\min_{y,t} \quad & \sum \nolimits_{i=1}^m c_it_i \tag{${\cal P}^{\rm ILP}_1$}\label{ILP1} \\
			{\rm{s.t.}}\quad & \sum \nolimits_{u:(u,v)\in E_{XX}\cup E_{UX}} y_{uv} = 1, \forall v\in X_L \label{C1} \\
			& \sum \nolimits_{v: (u,v)\in E_{XX}\cup E_{UX}} y_{uv} \le 1, \forall u \in X_R\cup U \label{C2}\\
			& \sum \nolimits_{j=1}^m w_{ij}t_j\ge 1, \forall i\in {\cal I} \label{C3} \\
			& t_j\ge \sum \nolimits_{(u,v)\in E_{u_j}}y_{uv}, \forall j\in {\cal U} \label{C3-add} \\
			& y_{uv}\in \{0,1\}, \forall (u,v)\in E_{XX}\cup E_{UX} \label{C4} \\
			& t_{j}\in \{0,1\}, \forall j\in {\cal U}. \label{C5}
		\end{align}
		%	Moreover, for an optimal solution $(y^\star, t^\star)$ to ${\cal P}_1^{\rm ILP}$, $S^\star=\{u_i: t^\star_i=1, i\in {\cal U}\}$ is an optimal solution to ${\cal P}_1$.
	\end{proposition}
	
	\begin{proof} Let $E_s=\{(u,v)\in E_{XX}\cup E_{UX}: y_{uv}=1, y \ {\rm subject\ to} \ (\ref{C1}), (\ref{C2}), (\ref{C4})\}$. Constraint (\ref{C1}) means every vertex of $X_L$ should be an end vertex of exactly one edge in $E_s$, and (\ref{C2}) means each vertex of $X_R\cup U$ can be the end vertex of at most one edge in $E_s$. Therefore, constraints (\ref{C1}), (\ref{C2}), and (\ref{C4}) make sure $E_s$ is a matching of ${\cal B}(A, B)$ that matches $X_L$. Moreover, constraint (\ref{C3}) means each source-SCC $X_i$ is input-reachable. Constraint (\ref{C3-add}) ensures if an edge of $E_{u_j}$ is included in the maximum matching $E_s$, then this input $u_j$ is selected (i.e., $t_j\ge 1$). By Lemma \ref{sc-theory}, any feasible solution $(y,t)$ subject to constraints (\ref{C1})-(\ref{C5}) corresponds to an input selection $S=\{u_i: t_i=1, i\in {\cal U}\}$ that makes the resulting system structurally controllable. It then follows immediately that ${\cal P}_1$ and ${\cal P}_{1}^{\rm ILP}$ are equivalent.
	\end{proof}
	
	\begin{proposition} \label{ILP-formulation} Under Assumption \ref{sc-assump}, ${\cal P}_2$ is equivalent to the following ILP ${\cal P}^{\rm ILP}_2$:
		\begin{align}
			\min_{y,t} \quad & \sum \nolimits_{i=1}^m c_it_i \tag{${\cal P}^{\rm ILP}_2$}\label{ILP2} \\
			{\rm{s.t.}}\quad & \sum \nolimits_{i=1}^m t_i \le k \ \label{C7} \\
			\quad & (\ref{C1}), (\ref{C2}), (\ref{C3}),  (\ref{C3-add}), (\ref{C4}), {\rm and}\ (\ref{C5}). \ \label{C8}
		\end{align}
		Again, any optimal solution $(y^\star, t^\star)$ to ${\cal P}_2^{\rm ILP}$ yields an optimal solution $S^\star=\{u_i: t^\star_i=1, i\in {\cal U}\}$ to ${\cal P}_2$.
	\end{proposition}
	
	\begin{proof}
		Note constraint (\ref{C7}) ensures the number of selected inputs is no more than $k$. Following a similar manner to the above analysis for ${\cal P}_1^{\rm ILP}$, the equivalence between ${\cal P}_2$ and ${\cal P}_{2}^{\rm ILP}$ is obtained.
	\end{proof}

	\begin{remark} It is notable that constraint (\ref{C2}) ensures that the right-hand side of (\ref{C3-add}) is at most $1$, implying that constraints (\ref{C3-add}) and (\ref{C5}) are compatible. Constraint (\ref{C3-add}) bridges conditions i) and ii) of Lemma \ref{sc-theory}.
	\end{remark}
	
	\begin{remark} From the proof of the equivalence between ${\cal P}_1$ and ${\cal P}_1^{\rm ILP}$, constraint (\ref{C8}) characterizes structural controllability. Hence, for a given $k\in {\mathbb N}$, ${\cal P}_2$ is feasible, if and only if ${\cal P}_2^{\rm ILP}$ is.
	\end{remark}

	%		In our ILP formulati
	\section{Polynomially Solvable ${\cal P}_1$ and ${\cal P}_2$} \label{main-result}

	This section provides conditions under which ${\cal P}_1$ and ${\cal P}_2$ are polynomially solvable. Our results show that, if the input structure satisfies certain `regularizations', which are characterized by the source-SCC-input incidence matrix $w$, irrespective of how the inputs connect vertices within each source-SCC or from the non-source-SCCs, then ${\cal P}_1$ and ${\cal P}_2$ can be solved in polynomial time via the corresponding LP-relaxations. 
	
	%We achieve this by showing with this condition, the constraint matrices of the respective ILPs are TU. We also discuss the solutions obtained via LP-rounding without this condition.
	
	\subsection{A polynomially solvable condition} %
	It has been shown in \cite{pequito2015complexity} ${\cal P}_1$ is NP-hard.  This fact is also reflected by constraint (\ref{C3}). Note provided that ${\cal B}(A)=(X_L,X_R,E_{XX})$ has a perfect matching, the constraints of ${\cal P}_1^{\rm ILP}$ reduce to constraints (\ref{C3}) and (\ref{C5}), which is the ILP formulation of the NP-hard set cover problem \cite{schrijver1998theory}.\footnote{Given a finite set ${\cal S}$ and a collection of its subsets $\{{\cal S}_1,...,{\cal S}_p\}$, the set cover problem is to select the minimum number of elements from $\{{\cal S}_1,...,{\cal S}_p\}$ such that their union is exactly ${\cal S}$.}
	A natural question is that, supposing for a certain specific class of systems associated with which optimizing $\sum \nolimits_{i=1}^m c_it_i$ subject to (\ref{C3}) and (\ref{C5}) can be implemented in polynomial time, can ${\cal P}_1$ (as well as ${\cal P}_2$) be solved efficiently?
	An already-known fact supporting the affirmative answer is that, if each available input is dedicated, then ${\cal P}_1$ is polynomially solvable. Can we broaden the class of systems on which ${\cal P}_1$ is polynomially solvable? %We shall give a positive answer here.
	
	To this end, we introduce the following constraint, named {\emph{source-SCC separated input constraint}}. We shall show, this constraint defines a large class of systems with which ${\cal P}_1$ and ${\cal P}_2$ are polynomially solvable.
	
	\begin{definition}(SSSI constraint) For $(A,B)$ in (\ref{plant}), it satisfies the SSSI constraint,  if no input vertices can connect with two different source-SCCs simultaneously in ${\cal G}(A,B)$. 	
	\end{definition}
	
	Note the SSSI constraint only requires that two different source-SCCs do not receive input signals from the same input. It does not impose any restrictions on how the inputs connect state vertices within each SCC (including the source-SCC and the non-source-SCC). Additionally, an input can simultaneously connect with one source-SCC and multiple non-source-SCCs. Hence, the SSSI constraint describes a wider class of input structures than the source-SCC grouped input constraint introduced in \cite{ACC2022-zhang} (the latter does not allow the existence of an input that simultaneously actuates a source-SCC and a non-source SCC).  {\emph {Particularly, the dedicated input structure is a special case of the SSSI constraint. A system that contains only one source-SCC automatically satisfies this constraint (a special case is that ${\cal G}(A)$ is strongly connected). }} See Fig. \ref{SSSI-example} for illustration of examples that satisfy the SSSI constraint.

	\begin{figure}
		\centering
		\includegraphics[width=2.4in]{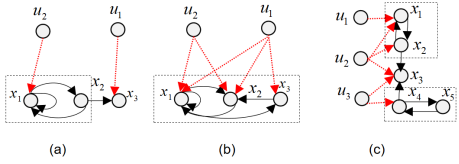}\\
		\caption{Examples of ${\cal G}(A,B)$ that satisfy the SSSI constraint: (a) dedicated input; (b) strongly connected case; (c) other case. Subgraphs in the boxes are the source-SCCs of ${\cal G}(A)$. }\label{SSSI-example}
	\end{figure}

	Our first main results are as follows. The proofs are postponed to the next subsection.
	
	\begin{theorem} \label{main-theo-1}
		Suppose $(A,B)$ satisfies Assumption \ref{sc-assump} and the SSSI constraint. Then, the following LP-relaxation ${\cal P}_1^{\rm LP}$ of ${\cal P}_1^{\rm ILP}$ always has an integral  optimal solution corresponding to the optimal solution of ${\cal P}_1$:
		\begin{align}
			\min_{y,t} \quad & \sum \nolimits_{i=1}^m c_it_i \tag{${\cal P}^{\rm LP}_1$}\label{LP1} \\
			{\rm{s.t.}}\quad & (\ref{C1}), (\ref{C2}), (\ref{C3}), {\rm and}\ (\ref{C3-add}) \label{CLP1} \\
			& 0\le y_{uv} \le 1, \forall (u,v)\in E_{XX}\cup E_{UX} \label{CLP4} \\
			& 0 \le t_{i} \le 1, \forall i\in {\cal U}. \label{CLP5}
		\end{align}
		Consequently, for $(A,B)$ satisfying the SSSI constraint, ${\cal P}_1$ can be solved in polynomial time.
	\end{theorem}
	
	\begin{theorem} \label{main-theo-2}
		Suppose $(A,B)$ satisfies Assumption \ref{sc-assump} and the SSSI constraint. Then, the following LP-relaxation ${\cal P}_2^{\rm LP}$ of ${\cal P}_2^{\rm ILP}$ always has an integral  optimal solution corresponding to the optimal solution of ${\cal P}_2$, whenever ${\cal P}_2$ is feasible.
		\begin{align}
			\min_{y,t} \quad & \sum \nolimits_{i=1}^m c_it_i \tag{${\cal P}^{\rm LP}_2$}\label{LP2} \\
			{\rm{s.t.}}\quad & (\ref{C1}), (\ref{C2}), (\ref{C3}), (\ref{C3-add}), (\ref{C7}), (\ref{CLP4}), {\rm and}\ (\ref{CLP5}). \label{CLP6}
		\end{align}
		Again, for $(A,B)$ satisfying the SSSI constraint, ${\cal P}_2$ can be solved in polynomial time.
	\end{theorem}
	
	In light of Theorems \ref{main-theo-1} and \ref{main-theo-2}, ${\cal P}_1$ and ${\cal P}_2$ can be solved in polynomial time via solving the respective LP-relaxations whenever the SSSI constraint is met.  Using off-the-shelf LP solvers, ${\cal P}_1^{\rm LP}$ and ${\cal P}_2^{\rm LP}$ can be solved in time $O((|E_{XX}\cup E_{UX}|+m)^{2.5}L)$ \cite{vaidya1989speeding}, where $|E_{XX}\cup E_{UX}|+m$ is the number of decision variables in those LPs, and $L=\log_2(c_{\max})+\log_2(k)+\log_2(n)$ is the number of input bits, with $c_{\max}\doteq\max_{1\le i \le m} c_i$ and $\{c_{i}\}$ being integral.\footnote{\label{foot2} When a non-integral optimal solution is found by an LP solver, an integral optimal solution can always be determined from it by computing the involved Hermite
		normal form in $\tilde O(d^w)$ time \cite{labahn2017fast}, where $d$ is the number of decision variables, $w<2.373$ is the exponent of matrix multiplication, and $\tilde O(\cdot)$ means logarithmic factors in the $O(\cdot)$
		are omitted; see \citep[Coro 5.3b, Theo 16.2]{schrijver1998theory} for details.} On the other hand, it seems unclear how to extend the graph-theoretic methods in \cite{A_Olshevsky_2015,pequito2016minimum} to the non-dedicated input case even with the SSSI constraint.   % It is easy to see that, without the SSSI constraint, ${\cal P}_1$ and ${\cal P}_2$ are at least as hard as the set cover problem. In this sense, it seems safe to say that, the SSSI
	%	constraint defines the `most possible' class of systems on which ${\cal P}_1$ and ${\cal P}_2$ permit polynomial time algorithms. Remarkably, as already mentioned, the SSSI constraint recovers two known polynomially solvable cases: the dedicated input case and the case with a strongly-connected ${\cal G}(A)$.
	
	The SSSI constraint defines a class of input structures where the source-SCCs may have certain autonomy (independence) so that they do not receive control signals/commands from the same input. In many practical network systems with geographically distributed subsystems, such as the power networks and ecological networks \cite{P.Fa2014Controllability,poggiale1998behavioural}, the dense interactions within subsystems correspond to strongly connected subgraphs. As subsystems are often geographically isolated, they cannot be directly affected by the same input. Hence, such network systems may satisfy the SSSI constraint. Other systems that may exhibit such an input structure may also be found in social networks, political
	networks, influence networks, etc. {For example, in political networks, the source-SCCs may correspond to different parties and the non-source-SCCs to voters without explicit partisans (non-party members), while the inputs correspond to the ideologies. It is often the case that different parties are influenced by distinct ideologies, leading to polarization \cite{ferreira2019modeling}, while the non-party members are more tolerant of different ideologies.} In social networks, the source-SCCs could represent groups that are separated by genders, families, countries, or even ideologies, such that different source-SCCs (serving as the decision groups) may not be influenced by the same input \cite{barabasi1999emergence,Y.Y.2011Controllability}. 
	
	%		 
	%	the decision group (corresponding to a source-SCC) of each party is in the charge of her party leaders, but is seldom influenced by other parties \cite{liu2012control}.
	
	\begin{remark}
		To our knowledge, ${\cal P}_2$ has seldom been considered before, and no prior work has reported polynomially solvable conditions for ${\cal P}_1$, except for the dedicated input case and the case that ${\cal G}(A)$ is strongly connected \cite{complexity_minimum_input,moothedath2018flow}. As for approximation algorithms, \cite{complexity_minimum_input} reduced ${\cal P}_1$ to the minimum set cover problem under the assumption that ${\cal B}(A)$ has a perfect matching.
		A flow-network based algorithm was proposed in \cite{moothedath2018flow} for ${\cal P}_1$, with the approximation factor equaling one plus the maximum number of source-SCCs that an input connects with simultaneously. When applied to systems satisfying the SSSI constraint, this algorithm only achieves a $2$-approximation factor without optimality guarantee.
	\end{remark}	
	
	\subsection{Analysis}  %{Proofs of Theorems \ref{main-theo-1} and \ref{main-theo-2}}
	This subsection gives the proofs of Theorems \ref{main-theo-1} and \ref{main-theo-2}. Our main idea is to prove that, the constraints matrices of ${\cal P}_1^{\rm ILP}$ and ${\cal P}_2^{\rm ILP}$ under the SSSI constraint are both TU.
	
	\begin{definition}[TU \cite{lawler2001combinatorial}] A matrix $M$ is TU if its every square submatrix has determinant $0, +1$, or $-1$.
	\end{definition}
	
	\begin{lemma}[\cite{hoffman201613}] \label{TU-integral}
		For a polyhedron $P=\{x\in {\mathbb{R}}^q: Mx\le b\}$, if $M$ is TU, then $P$ is integral (i.e., every vertex or extreme point of $P$ is integral) for any integral $b$.
	\end{lemma}
	
	%	It is known that for a polyhedron $P=\{x\in {\mathbb{R}}^q: Mx\le b\}$, if $M$ is TU, then $P$ is integral (i.e., every vertex or extreme point of $P$ is integral) for any integral $b$ \cite{hoffman201613}.
	According to the fundamental theorem of LP, every optimal solution of an LP (if exists) is either a vertex of its feasible polyhedron (i.e., feasible region), or lies on a face of optimal solutions (i.e., being a convex combination of its vertices that are the optimal solutions) \cite{boyd2004convex}. From Lemma \ref{TU-integral}, for an LP $\min \{c^\intercal x| Mx\le b, x\in {\mathbb{R}}^q\}$ with $M$ being TU, it always has integral optimal solutions for any integral $b$ and all rational $c$ whenever the optimum exists and is finite \cite{hoffman201613}. It immediately follows that the ILP $\min \{c^\intercal x| Mx\le b, x\in {\mathbb{Z}}^q\}$ with $M$ being TU can be solved efficiently by simply solving its corresponding LP-relaxation $\min \{c^\intercal x| Mx\le b, x\in {\mathbb{R}}^q\}$. Particularly, this implementation has polynomial time complexity for solving the original ILP (see \citep[Theo 16.2]{schrijver1998theory} and footnote \ref{foot2}). % for details on how to recover the integral optimal solutions when the corresponding LP-relaxation returns fractional optimal solutions
	
	Given $(A,B)$, let $n_E\doteq |E_{UX}\cup E_{XX}|$, $n_V\doteq|X_L\cup U\cup X_R|= 2n+m$. Rewrite $E_{XX}\cup E_{UX}=\{e_1,...,e_{n_E}\}$ and $X_L\cup U\cup X_R=\{v_1,...,v_{n_V}\}$. Associated with $(A,B)$ we construct two matrices $M\in \{0,\pm 1\}^{(2n+2m+r)\times (n_E+m)}$ and $\hat M \in \{0,\pm 1\}^{(2n+2m+r+1)\times (n_E+m)}$ for ${\cal P}_1^{\rm ILP}$ and ${\cal P}_2^{\rm ILP}$, respectively as follows:
	{\footnotesize
		\begin{equation} \label{M-eq}
			M_{ij}=\left\{ \begin{aligned}
				& 1, {\text {if}} \ v_i\in \partial(e_j), 1\le i \le n_V, 1\le j \le n_E \\
				& -w_{i-n_V,j-n_E}, {\text {if}} \ n_V+1\le i \le n_V+r, n_E+1\le j \le n_E+m \\
				& 1, {\text {if}} \ n_V+r+1\le i \le n_V+r+m, e_j\in E_{u_{i-n_V-r}} \\
				&-1, {\text {if}} \ n_V+r+1\le i \le n_V+r+m, j=n_E+i-n_V-r \\
				& 0, {\text {otherwise}},
			\end{aligned}\right.
	\end{equation} }
	\begin{equation} \label{M-hat-eq} \hat M=\left[\begin{array}{c}
			M	\\
			\alpha	
		\end{array}\right]	\end{equation}
	where $\partial(e_j)$ represents the vertices in edge $e_j$, and $\alpha\doteq [0_{1\times n_E},1_{1\times m}]$.
	It is clear that, in constructing $M$,  the first item corresponds to constraints (\ref{C1}) and (\ref{C2}) of ${\cal P}_1^{\rm ILP}$, the second item to constraint (\ref{C3}), while the third and fourth items to constraint (\ref{C3-add}) (see (\ref{LP-formal}) for the aggregated equation of those constraints). In addition, $\alpha$ of $\hat M$ corresponds to constraint (\ref{C7}).
	
	\begin{figure}
		\centering
		\includegraphics[width=1.4in]{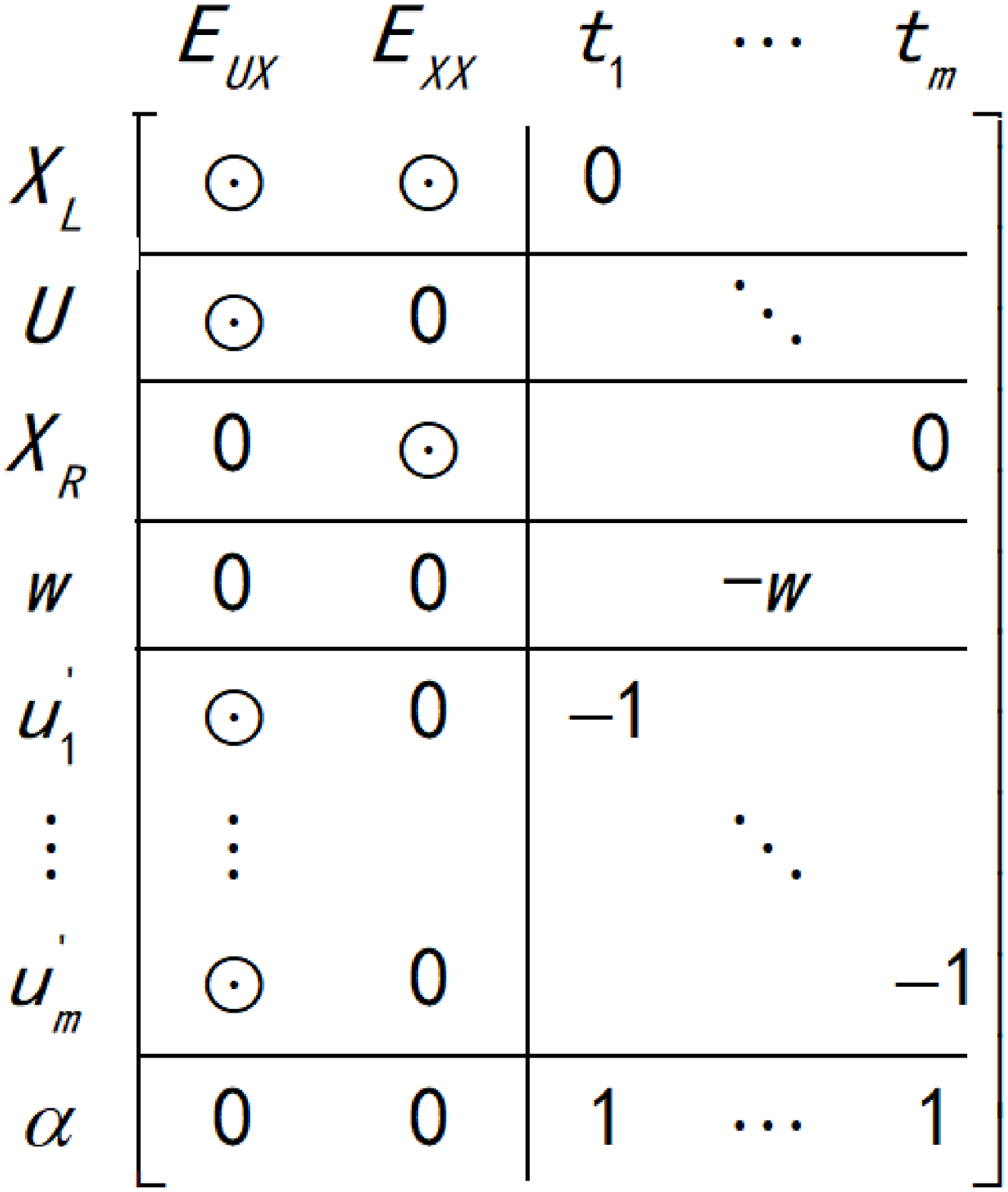}\\
		\caption{Illustration of $M$ and $\hat M$ in the proof of Proposition \ref{TU-M}. $\bigodot$ denotes a matrix block with entries from $\{0,1\}$. $X_L,...,\alpha$ ($E_{UX},...,t_m$) are the row (column) indices.}\label{proof-illustration}
	\end{figure}
	
	The following proposition characterizes the TU property of $M$ and $\hat M$ with the SSSI constraint, which is crucial to our results.
	
	\begin{proposition} \label{TU-M}
		Suppose $(A,B)$ satisfies the SSSI constraint. Then, both matrices $M$ and $\hat M$ are TU.
	\end{proposition}
	
	There are many different characterizations for TU matrices (see \citep[Chap 19]{schrijver1998theory}). Our proof relies on the following Ghouila-Houri's characterization of TU.
	
	\begin{lemma}\label{Ghouila-houri}(\citep[(iv) of Theorem 19.3]{schrijver1998theory}) A $p\times q$ integral matrix ${\cal A}=[a_{ij}]$ is TU, if and only if each set $R\subseteq \{1,...,p\}$ can be divided into two disjoint subsets $R_1$ and $R_2$ such that
		\begin{equation} \label{TU-sum}
			\sum \nolimits_{i\in R_1}a_{ij} -\sum \nolimits_{i\in R_2} a_{ij} \in \{-1,0,1\}, j=1,...,q. 	
		\end{equation}
	\end{lemma}
	
	{\bf{Proof of Proposition \ref{TU-M}}}:  {\bf We first prove the TU of $M$.} For the ease of description, suppose corresponding to the respective constraints of ${\cal P}_1^{\rm ILP}$, the rows of $M$ are indexed by $X_L$, $U$, $X_R$, $w$, and $u'_1,...,u'_m$, and columns are indexed by $E_{UX}, E_{XX}$, and $t_1,...,t_m$; see Fig. \ref{proof-illustration} for illustration. % For a submatrix $M'$ of $M$, by saying the rows (columns) of $M'$ indexed by $X_L$ ($U,X_R,\cdots $), we mean the rows (columns) that are both in $M'$ and corresponding to the maximal subset of $X_L$ ($U,X_R,\cdots $).
	We shall prove that every square $k\times k$ $(k\in {\mathbb N}_+)$ submatrix $M'$ of $M$ is TU by induction. For the beginning with $k=1$, $M'$ is certainly
	TU since each entry of $M$ is among $\{0,\pm 1\}$. Suppose this claim is true for all $(k-1)\times (k-1)$ submatrices ($k\ge 2$). Consider an arbitrary $k\times k$ submatrix $M'$ of $M$. If $M'$ contains a zero column, then $\det M'=0$. If $M'$ contains a column that has only one nonzero entry, then $\det M'=\pm \det M''\in \{0,\pm 1\}$, where $M''$ is the submatrix of $M'$ after deleting the respective row and column of that nonzero entry. Hence, we only need to consider the case where each column of $M'$ has at least two nonzero entries. This case will be divided into two subcases, detailed as follows.
	
	{\bf Subcase i:} $M'$ does not contain rows indexed by $u'_1,...,u'_m$ (corresponding to constraint (\ref{C3-add})). Since every column of $w=[w_{ij}]$ contains at most one nonzero entry with the SSSI constraint, $M'$ must consist of rows and columns indexed by subsets of $X_L\cup U\cup X_R$ and $E_{UX}\cup E_{XX}$, respectively. Notice that $(X_L\cup U \cup X_R, E_{UX}\cup E_{XX})$ is bipartite with bipartitions $X_L$ and $X_R\cup U$, and every column of $M'$ contains exactly $2$ nonzero entries. We can always partition the rows of $M'$ into two parts $R_1$ and $R_2$, such that each $R_i$ contains exactly one $1$ in each of its columns. Consequently, $\sum \nolimits_{i\in R_1}M'_{ij} -\sum \nolimits_{i\in R_2} M'_{ij}=0$ for each column of $M'$. By Lemma \ref{Ghouila-houri}, $M'$ is TU.

	{\bf Subcase ii:} $M'$ contains some rows indexed by subsets of $\{u'_1,...,u'_m\}$. Without harming generality, assume $M'$ contains rows indexed by $\{u'_1,...,u'_q\}$, $1\le q \le m$.
	Since each column indexed by $t_1,...,t_m$ contains at most two nonzero entries, $M'$ must contain rows indexed by the first $q$ rows of $w$ and columns indexed by $\{t_1,...,t_q\}$ (as otherwise there exists a column of $M'$ that does not have two nonzero entries).  Let us partition rows of $M'$ into disjoint sets $R_1,R_2,...,R_5$ from the top down, such that $R_1$ is a subset of $X_L$, $R_2$ is of $\{u_1,...,u_q\}$, $R_3$ is of $\{u_{q+1},...,u_{m}\}\cup X_R$, $R_4$ is of the first $q$ rows of $w$, and $R_5=\{u'_1,...,u'_q\}$ (note some sets may be empty). Suppose further in the {\emph{rows of $M'$ indexed by $R_5$ and columns indexed by $E_{UX}\cup E_{XX}$}}, the nonzero columns are indexed by $C_5$ (thus $C_5\subseteq E_{UX}$). The remaining columns of $M'$ are indexed by $\bar C_5$. With those partitions, it can be verified that for each column $j\in C_5$, {\small
		\begin{equation}\label{E_UX} \underbrace{\sum \limits_{i\in R_1} M'_{ij}+ \sum \limits_{i\in R_2} M'_{ij}}_{ 1 \ {\emph{or}} \ 2}- \underbrace{\sum \limits_{i\in R_3} M'_{ij}}_{0}+\underbrace{\sum \limits_{i\in R_4} M'_{ij}}_{0}-\underbrace{\sum \limits_{i\in R_5} M'_{ij}}_{1}=0,1,\end{equation}}which comes from the fact that $M'_{ij}=0$, $\forall i\in R_3\cup R_4$,  there is exactly one $i\in R_5$ with $M'_{ij}=1$ (by the definition of $R_5$ and recalling that each edge of $E_{UX}$ is incident to exactly one vertex of $X_L$ and one vertex of $U$ (or $\{u_1',...,u_m'\}$)), and that each column of $M'$ with rows indexed by $R_1\cup R_2$ and column by $j$ has at least one $1$ and at most two $1$'s.
	Similarly, for each column $j\in \bar C_5\cap (E_{UX}\cup E_{XX})$, {\small
		\begin{equation}\label{E_XX} \underbrace{\sum \limits_{i\in R_1} M'_{ij}}_{1}+ \underbrace{\sum \limits_{i\in R_2} M'_{ij}}_{0}- \underbrace{\sum \limits_{i\in R_3} M'_{ij}}_{1}+\underbrace{\sum \limits_{i\in R_4} M'_{ij}}_{0}-\underbrace{\sum \limits_{i\in R_5} M'_{ij}}_{0}=0,\end{equation}}which is due to the fact that, 
	there are exactly two $1$'s in the column of $M'$ indexed by $j$ and rows indexed by $R_1\cup R_3$, and $M'_{ij}=0$ for $i\in R_2\cup R_4 \cup R_5$ (note the row indexed by $u_i'$ with columns indexed by $E_{UX}\cup E_{XX}$ is the same as the row indexed by $u_i$ with columns indexed by $E_{UX}\cup E_{XX}$, $i=1,...,m$, leading to $M_{ij}'=0, \forall i\in R_2\cup R_5$). In addition, for each column $j\in \bar C_5 \cap  \{t_1,...,t_p\}$, {\small
		\begin{equation}\label{t1tm} \underbrace{\sum \limits_{i\in R_1} M'_{ij}}_{0}+ \underbrace{\sum \limits_{i\in R_2} M'_{ij}}_{0}- \underbrace{\sum \limits_{i\in R_3} M'_{ij}}_{0}+\underbrace{\sum \limits_{i\in R_4} M'_{ij}}_{-1}-\underbrace{\sum \limits_{i\in R_5} M'_{ij}}_{-1}=0,\end{equation}}which is because in the column of $M'$ indexed by $j$, there are exactly two $-1$'s in the rows indexed by $R_4\cup R_5$, and all the other rows are zeros. By Lemma \ref{Ghouila-houri}, $M'$ is TU.
	
	Hence, by induction, we conclude that $M$ is TU.

	{\bf We now prove the TU of $\hat M$.} We still do this by induction. For the beginning, every $1\times 1$ submatrix of $\hat M$ is certainly TU. Assume that every $(k-1)\times (k-1)$ submatrix of $\hat M$ is TU ($k\ge 2$). Let $\hat M'$ be a $k\times k$ submatrix of $\hat M$. Similar to the above analysis,  we only need to show $\hat M'$ is TU subject to the constraint that each of its columns has at least two nonzero entries. Since we have proven $M$ is TU, it suffices to show each $\hat M'$ that contains elements from the last row $\alpha$ of $\hat M$ is TU.  From the above analysis in {\bf subcase ii}, if every column of $\hat M'$ indexed by $E_{UX}\cup E_{XX}$ contains at least two nonzero entries, then there is an assignment of signs for rows of $\hat M'$ with columns indexed by subsets of $E_{UX}\cup E_{XX}$, such that their sum is a row vector with entries in $\{0,1\}$, in which the rows indexed by the subset of $\{u_1',...,u_m'\}$ (i.e., $R_5$) have sign $-1$. Moreover,  for the columns of $\hat M'$ indexed by a subset of $\{t_1,...,t_m\}$, let us assign $-1$'s to the signs of rows corresponding to subsets of $\{u_1',...,u_m'\}$ and $w$, as well as $\alpha$. Then, the sum of those signed rows is a vector with entries in $\{0,1\}$. This is because each entry is the sum of exactly one $-1$ and at least one $+1$ (at most two $+1$'s). Hence, by Lemma \ref{Ghouila-houri}, $\hat M'$ is TU. By induction, we know $\hat M$ is TU. \hfill $\square$
	
	%Notice that for each column $j$ of $\hat M'$ indexed by $\{t_1,...,t_m\}$,
	%$$\sum \limits_{i\in \{\}} M'_{ij}+ \sum \limits_{i\in R_2} M'_{ij}- \sum \limits_{i\in R_3} M'_{ij}-\sum \limits_{i\in R_4} M'_{ij}-\sum \limits_{i\in R_5} M'_{ij}=0 $$
	
	We are now proving Theormes \ref{main-theo-1} and \ref{main-theo-2}.
	
	{\bf Proof of Theorem \ref{main-theo-1}:} As analyzed above, it suffices to prove that the constraint matrix of ${\cal P}_1^{\rm LP}$ is TU. To this end, rewrite the constraints of ${\cal P}_1^{\rm LP}$ as
	\begin{equation} \label{LP-formal}
		\underbrace{\left[\begin{array}{l}
				M\\
				-M_{eq}\\
				I_{n_E+m}\\
				-I_{n_E+m}
			\end{array}\right]}_{M_{LP}}\left[\begin{array}{l}
			y\\
			t
		\end{array}\right]\le \left[\begin{array}{l}
			1_{n_V\times 1}\\
			-1_{r\times 1}\\
			0_{m\times 1}\\
			-1_{n\times 1}\\
			1_{(n_E+m)\times 1} \\
			0_{(n_E+m)\times 1}
		\end{array}\right],
	\end{equation}
	where $M_{eq}$ consists of rows of $M$ corresponding to constraint (\ref{C1}).
	Since $M$ is TU from Proposition \ref{TU-M}, upon defining  $M'_{LP}\doteq {\tiny {\left[\begin{array}{l}
				\ \ \ \ M\\
				I_{n_E+m}
			\end{array}\right]}}$, $M'_{LP}$ is also TU. This is because, any square submatrix $M'$ that contains elements from the last $n_E+m$ rows of $M_{LP}'$ must have a determinant $\pm \det M''\in \{0,\pm 1\}$, where
	$M''$ is the submatrix of $M'$ after deleting the respective rows and columns of the elements in the last $n_E+m$ rows of $M_{LP}'$.  As $M_{LP}$ is obtained from $M'_{LP}$ by duplicating its rows (with negative signs),
	$M_{LP}$ is certainly TU by definition. The required statement follows directly from the TU of $M_{LP}$.  \hfill $\square$
	
	{\bf Proof of Theorem \ref{main-theo-2}:} Again, it suffices to show that the constraint matrix of ${\cal P}_2^{\rm LP}$ is TU.
	Since $\hat M$ is TU, this can be done similarly to the proof of Theorem \ref{main-theo-1}. Details are omitted due to their similarities.  \hfill $\square$
	
}

\subsection{Generalization of SSSI constraint}
In this subsection, we extend the SSSI constraint to a general algebraic condition on the source-SCC-input incidence matrix $w$, namely, the restricted TU introduced in this paper, which allows each input to connect with multiple source-SCCs while preserving the polynomial solvability. 

%The source-SCC-input incidence matrix $w$ satisfies the restricted TU condition:

\begin{definition}\label{regular-cond}(Restricted TU) The matrix $w\in \{0,1\}^{r\times m}$ is said to be restrictedly TU, if
$\left[\begin{array}{l}
	\quad	w \\
	1_{1\times m}
\end{array}\right]$ is TU. Equivalently, each set $R\subseteq \{1,...,r\}$ can be divided into two disjoint subsets $R_1$ and $R_2$, such that
\begin{equation} \label{TU-sum-restricted}
	\sum \nolimits_{i\in R_1}w_{ij} -\sum \nolimits_{i\in R_2} w_{ij} \in \{0,1\}, j=1,...,m.
\end{equation}
\end{definition}

\begin{remark}
	The equivalence between the two conditions in Definition \ref{regular-cond} results from Lemma \ref{Ghouila-houri}. Say, for $w$ to be TU, any sub-rows of $w$ should satisfy (\ref{TU-sum}). For $[w^\intercal, 1_{m\times 1}]^\intercal$ to be TU, any sub-rows of $w$ with the addition of $1_{1\times m}$ should also satisfy (\ref{TU-sum}). This leads to that any sub-rows of $w$ need to satisfy (\ref{TU-sum-restricted}).
\end{remark}

\begin{theorem} \label{extend-theorem}
	Suppose $(A,B)$ satisfies Assumption \ref{sc-assump} and the associated $w$ is restrictedly TU. Then, the LP-relaxations ${\cal P}_1^{\rm LP}$ and ${\cal P}_2^{\rm LP}$ both have integral optimal solutions. That is, ${\cal P}_1$ and ${\cal P}_2$ then can be solved in polynomial time via solving the respective ${\cal P}_1^{\rm LP}$ and ${\cal P}_2^{\rm LP}$.
\end{theorem}

\begin{proof}
	Similar to the proof of Theorems \ref{main-theo-1} and \ref{main-theo-2}, it suffices to show that $M$ and $\hat M$ defined in (\ref{M-eq}) and (\ref{M-hat-eq}) are both TU under the restricted TU of $w$.
	Again, we achieve this by showing that every $k\times k$ submatrix of $M$ is TU via induction for $k\in {\mathbb N}_+$. The beginning case with $k=1$ is certainly true. Now assume the claim holds for some $k-1$ ($k\ge 2$). Let $M'$ be a $k\times k$ submatrix of $M$. Based on the proof of Proposition \ref{TU-M}, it is enough to consider the case where each column of $M'$ contains at least two nonzero entries.  Following the analysis in {\bf subcase ii}, let us change $R_4$ to the indices indexing the rows of $w$ that are contained in $M'$, and definitions of the remaining $R_1, R_2, R_3, R_5, C_5$ and $\bar C_5$ remain unchanged. For each column $j$ of $M'$ indexed by an element of $E_{UX}\cup E_{XX}$, it is obvious that the equalities (\ref{E_UX}) and (\ref{E_XX}) hold, since $M'_{ij}\equiv 0$ $\forall i\in R_4$. For each column $j\in {\bar C}_5\cap \{t_1,...,t_m\}$ of $M'$, by Definition \ref{regular-cond}, $R_4$ can be partitioned into $R_4=R_{41}\cup R_{42}$, such that
	\begin{equation} \label{sub-sum}
		\sum \nolimits_{i\in R_{41}}M'_{ij} -\sum \nolimits_{i\in R_{42}} M'_{ij} \in \{0,-1\}.
	\end{equation}
	By changing $\sum \nolimits_{i\in R_4} M'_{ij}$ in (\ref{t1tm}) to the left-hand side of (\ref{sub-sum}), we get that the right-hand side of (\ref{t1tm}) is in $\{\pm 1, 0\}$ ($-1$ could appear because it may happen that $M'_{ij}=0$ $\forall i\in R_5$). Consequently, $M'$ is TU by Lemma \ref{Ghouila-houri}. This leads to the TU of $M$.
	
	We are to show $\hat M$ is TU.  We still resort to induction. Suppose every $(k-1)\times (k-1)$ submatrix of $\hat M$ is TU for some $k\ge 2$. To demonstrate the case with $k$, since $M$ is TU as proved above,  it suffices to prove that every $k\times k$ submatrix of $\hat M$ that contains sub-columns of the last row $\alpha$ is TU. Denote such a matrix by $\hat M'$. Similarly, only the case that each column of $\hat M'$ contains at least two nonzero entries needs to be considered. For the submatrix of $\hat M'$ obtained by removing its last row, let $R_1,R_2,...,C_5,\bar C_5$ be defined in the same way as in the proof for $M$ under the restricted TU condition. And by assumption, there is a partition $R_4=R_{41}\cup R_{42}$ such that (\ref{sub-sum}) holds. We declare that, for each column $j$ of $\hat M'$, it holds
	{\small{\begin{equation}\label{extended-subsum} \begin{array}{l}
					\sum \limits_{i\in R_1} \hat M'_{ij}+ \sum \limits_{i\in R_2} \hat M'_{ij}-\sum \limits_{i\in R_3} \hat M'_{ij}-\left(\sum \limits_{i\in R_{41}}\hat M'_{ij} -\sum \limits_{i\in R_{42}} \hat M'_{ij}\right) \\
					-\sum \limits_{i\in R_5} \hat M'_{ij}-\alpha_{j}\in \{0,\pm 1\}
				\end{array}.\end{equation}}}Indeed, for each column $j$ of $\hat M'$ indexed by an element of $E_{UX}\cup E_{XX}$,  (\ref{extended-subsum}) holds because of the same reasoning as (\ref{E_UX}) and (\ref{E_XX}), noting $\alpha_j=0$. And for each column $j$ of $\hat M'$ indexed by a subset of $\{t_1,...,t_m\}$, due to (\ref{sub-sum}) and $\alpha_j=1$, the left-hand side of (\ref{extended-subsum}) is the sum of exactly one $-1$, and at most two $1$'s. Hence, the $k\times k$ submatrix $\hat M'$ is TU, which indicates $\hat M$ is TU by induction.
\end{proof}

Theorem \ref{extend-theorem} reveals, although ${\cal P}_1$ and ${\cal P}_2$ are NP-hard in general, provided that the source-SCC-input incidence matrix satisfies certain `regularizations', irrespective of how each input connects vertices within the same source-SCC or vertices belonging to (different) non-source SCCs,  those problems can be solved in polynomial time. According to \citep[Theo 20.3]{schrijver1998theory}, the TU of a given $m\times n$ matrix can be tested in time $O((m+n)^4m)$. This means the restricted TU of $w$ can be verified in time $O((r+m)^4r)$. Hence, in practice, one could first check the restricted TU of $w$. If the answer is yes, then the {\emph{optimal}} solutions to the associated ${\cal P}_1$ and ${\cal P}_2$ can be determined efficiently.

TU matrices can be fully characterized using the so-called network matrices combined with some basic operations (see \citep[Theo 19.6]{schrijver1998theory}), which is closely related to certain graphical structures (\citep[Page 276]{schrijver1998theory}). This means, the restricted TU on $w$ may correspond to some graphical characterizations of the input structure, which is left for future work.
In the following, in addition to the SSSI constraint (which certainly satisfies the restricted TU condition, since the sum of any sub-rows of $w$ is a row vector with entries in $\{0,1\}$),  we provide some extra easily-verified scenarios where the restricted TU condition is met (see Example \ref{restric-TU-examp} for illustrations), whose proofs are postponed to the appendix: 
\begin{itemize}
	\item Extended SSSI constraint:    The $r$ source-SCCs could be partitioned into $l$ disjoint groups, with their indices being $\{C_i|_{i=1}^l\}$, $C_i\subseteq {\cal I}$, and $\bigcup \nolimits_{i=1}^l C_i={\cal I}$. Each member in the same group has the same `input configuration', that is, the inputs that connect with each source-SCC in the same group are the same.  Mathematically, for the matrix $w$ defined in Section \ref{main-ILP}, upon letting $S_{ik}=\{j\in {\cal U}: w_{kj}=1\}$ for each $i\in \{1,...,l\}$ and $k\in C_i$, we have $S_{ik_1}=S_{ik_2}$ for any $k_1,k_2\in C_i$, $\forall i$. It is easy to see, the SSSI constraint corresponds to that $|C_i|=1$, $i=1,...,l$. Additionally, the case that $B$ is a {\emph{full matrix}} with all entries being nonzero also satisfies this constraint, with $l=1$, $C_1={\cal I}$, and $S_{1k}={\cal U}$ for each $k\in {\cal I}$.

	\item Row-monotone (column-monotone) constraint:
	We say a row (column) vector $a=(a_1,\cdots, a_p)$ is non-decreasing, if $a_1\le a_2 \le \cdots \le a_p$, and non-increasing if $a_1\ge a_2 \ge \cdots \ge a_p$.
	The source-SCC-input incidence matrix $w$ is said to satisfy the row-monotone (resp. column-monotone) constraint, if all of its rows (resp. columns) are either non-increasing or non-decreasing.

	\item Permutable row/column-monotone constraint:  $w$ is said to satisfy the permutable row/column-monotone constraint, if after some row and column permutations,
	the obtained $w$ satisfies the row/column-monotone constraint.
\end{itemize}

Note all the above input constraints allow that, one input can actuate multiple state vertices belonging to {\emph{different}} source-SCCs.  Further, if $w$ is {\emph{block-diagonal}} with each diagonal block satisfying the (permutable) row/column-monotone constraints, then $w$ is restrictedly TU.

\begin{example} \label{restric-TU-examp}
	We provide some examples of $w$ that satisfy the constraints mentioned above. Consider matrices
	$$w_1=\left[\begin{array}{cccccc}
		1 &  1 & 0 & 0 & 0 & 0 \\
		1 &  1 & 0 & 0 & 0 & 0 \\
		0 & 0 & 1 & 1 & 1 & 0 \\
		0 & 0 & 1 & 1 & 1 & 0 \\
		0 & 0 & 0 & 0 & 0 & 1
	\end{array}\right],w_2=\left[\begin{array}{cccc}
		1 &  1 & 1 & 0 \\
		1 &  0 & 1 & 0\\
		1 & 0 & 0 & 0
	\end{array}\right],$$
	$$w_3=\left[\begin{array}{cccc}
		1 &  1 & 1 & 0 \\
		0 &  1 & 0 & 0\\
		0 & 1 & 1 & 0
	\end{array}\right],w_4=\left[\begin{array}{ccc}
		1 & 0 & 0 \\
		0 &  1 & 0 \\
		1 &  0 & 1 
	\end{array}\right].$$
	It is easy to see, $w_1, w_2$, and $w_3$ respectively satisfy the extended SSSI constraint, the column-monotone constraint, and the permutable column-monotone constraint. After some row and column permutations, $w_4$ is turned into a block-diagonal matrix with each diagonal block being row-monotone. Therefore, they are all restrictedly TU. \hfill $\square$
\end{example} 

\begin{figure}
	\centering
	\includegraphics[width=1.68in]{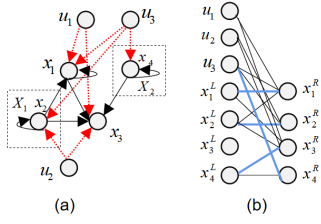}\\
	\caption{System digraph ${\cal G}(A,B)$ (a) and the associated bipartite graph ${\cal B}(A,B)$ (b) (borrowed from \cite{moothedath2018flow}). Bold blue edges in (b) constitute a maximum matching of ${\cal B}(A,B)$.} \label{restricted-TU-example}
\end{figure}

\begin{example}
	Consider the system $(A,B)$ in \cite{moothedath2018flow},  with the associated  ${\cal G}(A,B)$ and ${\cal B}(A,B)$ given respectively in Figs. \ref{restricted-TU-example}(a) and (b). From Fig. \ref{restricted-TU-example}(a), ${\cal G}(A,B)$ contains two source-SCCs, with their vertex sets being $X_1=\{x_2\}$ and $X_2=\{x_4\}$. The corresponding source-SCC-input matrix $w$ is
	$$w=\left[\begin{array}{ccc}
		0 &  1 & 1 \\
		0 &  0 & 1 
	\end{array}\right].$$
	Since $w$ is row-monotone, it is certainly restrictedly TU. Indeed, given any positive cost to each input, it turns out that the ${\cal P}_1^{\rm LP}$ and ${\cal P}_2^{\rm LP}$ always have an integral optimal solution $t^\star=[0,0,1]$, implying the optimal solutions to ${\cal P}_1$ and ${\cal P}_2$ of this system are $S^\star=\{u_3\}$.   \hfill $\square$
\end{example}

%\begin{remark}
%	We remark that if $w$ is TU but not restrictedly TU, $M$ may not be TU. This can be shown from the following example. 
%\end{remark}

\section{Solutions via LP-relaxation and LP-rounding: general case} \label{general-case}
In this section, we discuss solutions to ${\cal P}_1$ obtained via the LP-relaxation and LP-rounding methods in the general case, i.e., without any restriction on $w$.
Particularly, we show that LP-relaxation can provide a tighter lower bound for ${\cal P}_1$ than the one obtained via the graph-theoretic method. The LP-rounding method, on the other hand, provides an upper bound for ${\cal P}_1$, which possesses provable approximation guarantee for a special case. 

% Particularly, these two methods give a lower and an upper bound for ${\cal P}_1$.

%Particularly, we show the LP-relaxation can provide a tighter lower bound for ${\cal P}_1$ than the one obtained via the graph-theoretic method. The LP-rounding method, on the other hand, provides an upper bound for ${\cal P}_1$, which possesses provable approximation guarantee for a special case. 

Let $c^{\star}_{\rm mat}$ be the minimum cost of input vertices that are matched in a maximum matching of ${\cal B}(A,B)$. By Lemma \ref{sc-theory}, $c^{\star}_{\rm mat}$, which can be obtained via the weighted maximum matching algorithms in polynomial time, is a lower bound for ${\cal P}_1$. In the following, we show that LP-relaxation can provide a tighter lower bound for ${\cal P}_1$ than $c^{\star}_{\rm mat}$.  %the one obtained via the aforementioned graph theoretic method

\begin{proposition} \label{prop-relax-1}
	Suppose $(A,B)$ satisfies Assumption \ref{sc-assump}. Let $c^{\star}_{\rm LP}$ be the optimal value of ${\cal P}_{1}^{\rm LP}$. Then $c^{\star}_{\rm LP}$ is a lower bound of the optimal value of ${\cal P}_1$ satisfying 
	$c^{\star}_{\rm LP}\ge c^{\star}_{\rm mat}$.
\end{proposition}

\begin{proof}
	It is obvious that $c^{\star}_{\rm LP}$ is a lower bound for ${\cal P}_1$ since the integer constraint in ${\cal P}_1^{\rm ILP}$ is dropped. Next, it is shown that $c^\star_{\rm mat}$ is the optimal value of the following problem 
	\begin{align}
		\min_{y,t} \quad & \sum \nolimits_{i=1}^m c_it_i  \tag{${\cal P}_1^{\rm RMA}$} \label{LP2add} \\
		{\rm{s.t.}}\quad & (\ref{C1}), (\ref{C2}), (\ref{C3-add}), (\ref{CLP4}), {\rm and}\ (\ref{CLP5}). %\label{CLP6add}
	\end{align}
	Indeed, to minimize $\sum \nolimits_{i=1}^m c_it_i$ subject to (\ref{C3-add}), it must hold $t_j=\sum \nolimits_{(u,v)\in E_{u_j}}y_{uv}$, $\forall j\in {\cal U}$. Hence, ${\cal P}_1^{\rm RMA}$ is equivalent to 
	\begin{align}
		\min_{y} \quad & \sum \nolimits_{i=1}^m c_i \sum \nolimits_{(u,v)\in E_{u_i}}y_{uv} \tag{${\cal P}_1^{\rm MAT}$} \\
		{\rm{s.t.}}\quad & (\ref{C1}), (\ref{C2}), (\ref{CLP4}), {\rm and}\ (\ref{CLP5}).
	\end{align}Since the constraint matrix of ${\cal P}_1^{\rm MAT}$ is TU (Proposition \ref{TU-M}), ${\cal P}_1^{\rm MAT}$ has an integral optimal solution which corresponds to the maximum matching of ${\cal B}(A,B)$ with the minimum cost for the inputs (notice an integral solution means at most one edge from an input is involved). Note the optimum of ${\cal P}_1^{\rm RMA}$ is a lower bound for that of ${\cal P}_1^{\rm LP}$ as constraint (\ref{C3}) is removed, proving that $c^{\star}_{\rm LP}\ge c^{\star}_{\rm mat}$.
\end{proof}

In what follows, we show a simple rounding technique can find feasible solutions to ${\cal P}_1$. And, this gives a $f$-approximation for ${\cal P}_1$ under the condition that ${\cal B}(A)$ has a perfect matching, with $f$ defined in the following: \begin{equation} \label{def1}
	f=\max_{i\in {\mathcal I}} \sum \nolimits_{j=1}^m w_{ij}.
\end{equation} In other words, $f$ is the maximum number of inputs that connect with a source-SCC simultaneously.

\begin{proposition} \label{prop-relax} Suppose $(A,B)$ satisfies Assumption \ref{sc-assump}. Let $(y^\star, t^\star)$ be an optimal solution to ${\cal P}_{1}^{\rm LP}$.
	Then, $S_{\rm int}=\{u_i: t_i^\star>0, i\in {\cal U}\}$ is a feasible solution to ${\cal P}_1$. Moreover, if ${\cal B}(A)$ has a perfect matching, then $S_{\rm int}$ gives a factor $f$-approximation for ${\cal P}_1$.
\end{proposition}

\begin{proof}
	We first prove $S_{\rm int}$ is feasible for ${\cal P}_1$. Let $t'$ be obtained by setting $t'_i=1$ if $t^{\star}_i>0$ and otherwise $t'_i=0$. It is then obvious that constraint (\ref{C3}) is met for $t'$. Moreover, $y^\star$ is a feasible point of the region subject to the constraints (\ref{C1}), (\ref{C2}), (\ref{C3-add}), (\ref{CLP4}), and (\ref{CLP5}), where $t_j$ in (\ref{C3-add}) should be replaced with $t_j'$ for each $j\in {\cal U}$ (denote this region by ${\cal R}$). According to Proposition \ref{TU-M}, we know readily that the constraint matrix of ${\cal R}$ is TU. Since $\cal R$ contains a feasible point $y^\star$, it must contain integral extreme points for the variable $y$,  and denote one of them by $y'$. Then, $(y',t')$ is a feasible integral solution to ${\cal P}_1^{\rm LP}$. It yields that $S_{\rm int}$ is feasible for ${\cal P}_1$.
	
	We now prove the second claim using the primal-dual argument. Since ${\cal B}(A)$ has a perfect matching, $t^\star$ must also be the optimal solution to the following problem
	\begin{align}
		\min_{t} \quad & \sum \nolimits_{i=1}^m c_it_i \tag{${\cal P}^{\rm LP'}_1$}\label{ILP-proof} \\
		{\rm{s.t.}}\quad	& \sum \nolimits_{j=1}^m w_{ij}t_j\ge 1, \forall i\in {\cal I} \label{C3-proof} \\
		& t_{i}\ge 0, \forall i\in {\cal U}. \label{C5-proof}
	\end{align}
	Indeed, if this is not true, then one can always find an optimal solution to ${\cal P}^{\rm LP'}_1$, which is feasible for ${\cal P}_1^{\rm LP}$ as constraints (\ref{C1}), (\ref{C2}), and (\ref{C3-add}) are met with $\sum \nolimits_{(u,v)\in E_{u_j}}y_{uv}=0$, $\forall j\in {\cal U}$, by the existence of a perfect matching of ${\cal B}(A)$. On the other hand, any optimal solution to ${\cal P}_1^{\rm LP'}$ gives a lower bound for the objective value of ${\cal P}_1^{\rm LP}$.
	
	Denote the optimal objective value of ${\cal P}_1$ by $c_{\rm opt}$. Then clearly $\sum \nolimits_{j=1}^m c_jt_j^\star\le c_{\rm opt}$. The dual LP of ${\cal P}^{\rm LP'}_1$ is
	\begin{align}
		\max_{z} \quad & \sum \nolimits_{i=1}^r z_i \tag{${\cal P}^{\rm LPD}_1$}\label{ILP-proof-dual} \\
		{\rm{s.t.}}\quad	& \sum \nolimits_{i=1}^r w_{ij}z_i\le c_j, \forall j\in {\cal U}  \\
		& z_{i}\ge 0, \forall i\in {\cal I}.
	\end{align}
	Let $z^\star$ be an optimal solution to ${\cal P}^{\rm LPD}_1$. By the strong duality of LP (see \citep[Chap 5.2.3]{boyd2004convex}), we have
	$$\sum \nolimits_{j=1}^m c_jt_j^\star= \sum \nolimits_{i=1}^r z_i^\star.$$
	The primal complementary slackness condition (see \citep[Chap 5.5]{boyd2004convex}) yields
	$${\rm whenever} \ t_j^\star \ne 0 \Rightarrow \sum \nolimits_{i=1}^r w_{ij}z_i^\star=c_j. $$
	Hence, we obtain
	{\small
	\begin{align*}
		\sum \nolimits_{j=1}^m c_jt_j' & = \sum \nolimits_{j: t_j^\star \ne 0} c_j  %\tag{${\cal P}^{\rm LPD}_1$}\label{ILP-proof-dual}
		= \sum \nolimits_{j: t_j^\star \ne 0} (\sum \nolimits_{i=1}^r w_{ij}z_i^\star) \\
		& =\sum \nolimits_{i=1}^r  \sum \nolimits_{j: t_j^\star \ne 0} w_{ij}z_i^\star 
		 \le \sum \nolimits_{i=1}^r  \sum \nolimits_{j=1}^m w_{ij}z_i^\star \\
		& \le \sum \nolimits_{i=1}^r  fz_i^\star 
		 = f \sum \nolimits_{j=1}^m c_jt_j^\star \\
		& \le f*c_{\rm opt},
	\end{align*}}where the last second inequality is due to (\ref{def1}), and the last equality to the strong duality.
	This ends the proof.
\end{proof} %\label{important-reason} 

\begin{remark}
	Proposition \ref{prop-relax} indicates we can simply pick all the nonzero entries in an optimal solution to ${\cal P}_1^{\rm LP}$ to get a feasible solution to ${\cal P}_1$. It is worth mentioning that a feasible solution can also be obtained via some graph-theoretic algorithms. Currently, we are not able to give an approximation bound for this LP-rounding based algorithm without the perfect matching condition. Combining Propositions \ref{prop-relax-1} and \ref{prop-relax} yields a lower and an upper bound for ${\cal P}_1$ (recalling the lower bound is exact if $w$ is restrictedly TU).  
\end{remark}

\section{Switched system case} \label{switched-case}
In this section, we extend results in the previous sections to the switched systems, focusing on the polynomially solvable conditions of ${\cal P}_4$ and ${\cal P}_5$.  First, ILP formulations of 
${\cal P}_4$ and ${\cal P}_5$ are given. Then, a joint SSSI constraint is proposed, under which it is shown ${\cal P}_4$ and ${\cal P}_5$ can be solved by the corresponding LP-relaxations. The restricted TU condition is also extended. Problems of selecting fixed inputs during the switching to achieve structural controllability are finally addressed and shown to be polynomially sovlable under the restricted TU condition.  
%\begin{lemma} 
%	$(A_{\sigma(\cdot)}, B)$ is controllable, if and only if $(A_{\sigma(\cdot)}, B_{\sigma(\cdot)})$ is controllable, with $B_1=B, B_2=\cdots = B_{p}=0_{n\times m}$. ($p$ is the number of switching modes)
%\end{lemma}	

For a structured matrix $M_1\in \{0,*\}^{n_1\times n_2}$, its generic rank, denoted as ${\rm grank}(M_1)$, is defined to be the maximum rank it can achieve as a function of its free parameters. 
For two structured matrices $M_1,M_2\in \{0,*\}^{n_1\times n_2}$, $M_3=M_1\vee M_2$ is a $n_1\times n_2$ structured matrix satisfying $M_{3,ij}=*$ if $M_{1,ij}=*$ or $M_{2,ij}=*$, otherwise $M_{3,ij}=0$. Define $\hat A = A_1\vee \cdots \vee A_p$, and $\hat B=[B_1,\cdots, B_p]$. Let ${\cal G}(\hat A)=(X,E_{\hat A})$ and ${\cal G}(\hat A, \hat B)=(X\cup \hat U, E_{\hat A}\cup E_{\hat B})$ be defined in the same way as in Section \ref{preliminary}, i.e.,  $X=\{x_1,...,x_n\}$, $\hat U=\{u_{11},..., u_{1m_1},..., u_{p1},..., u_{pm_p}\}$,
$E_{\hat A}=\{(x_j,x_i): \hat A_{ij}\ne 0\}$, and $E_{\hat B}=\{(u_{ki},x_j): B_{k,ji}\ne 0\}$.
With these notations, the following lemma characterizes the structural controllability of the switched system (\ref{plant-switched}). 

\begin{lemma} \cite{LiuStructural} \label{switch-lemma}
	For system (\ref{plant-switched}),	$(A_{\sigma(\cdot)}, B_{\sigma(\cdot)})$ is structurally controllable, if and only if ~\\ i) ${\rm grank}([A_1,\cdots, A_p, B_1,\cdots, B_p])=n$, and ii) every state vertex of $X$ in ${\cal G}(\hat A, \hat B)$ is input-reachable.  %B_1\vee\cdots \vee B_p
\end{lemma}	

%Based on Lemma \ref{switch-lemma}, define $\hat A = A_1\vee \cdots \vee A_p$, and $\hat B=[B_1,\cdots, B_p]$. Let ${\cal G}(\hat A)=(X,E_{\hat A})$ and ${\cal G}(\hat A, \hat B)=(X\cup \hat U, E_{\hat A}\cup E_{\hat B})$ be defined in the same way as in Section \ref{preliminary}, i.e.,  $X=\{x_1,...,x_n\}$, $\hat U=\{u_{11},..., u_{1m_1},..., u_{p1},..., u_{pm_p}\}$,
%$E_{\hat A}=\{(x_j,x_i): \hat A_{ij}\ne 0\}$, and $E_{\hat B}=\{(u_{ki},x_j): B_{k,ji}\ne 0\}$. ii) every sate node in ${\cal G}(A_1\vee\cdots\vee A_p, [B_1,\cdots, B_p])$ is input-reachable.  %B_1\vee\cdots \vee B_p

Decompose ${\cal G}(\hat A)$ into SCCs, and suppose there are $\hat r$ source-SCCs, and let 
$\hat {\cal I}=\{1,...,\hat r\}$ be their indices. For the $k$th mode, $1\le k \le p$, the source-SCC-input incidence matrix $w^k\in \{0,1\}^{\hat r\times m_k}$ is defined as $w^k_{ij}=1$ if $u_{kj}$ connects with the $i$th source-SCC, otherwise 
$w^k_{ij}=0$. Furthermore, let $\hat A'=[A_1,..., A_p]$. Associated with $[\hat A', \hat B]$, define the bipartite graph ${\cal B}(\hat A', \hat B)=(X, \hat X \cup \hat U, E_{\hat X X}\cup E_{\hat U X})$, where the vertex set $\hat X=\{x_{11},..., x_{1n},..., x_{p1},..., x_{pn}\}$, $E_{\hat X X}=\{(x_{kj},x_i): A_{k,ij}\ne 0, k=1,..., p\}$, and $E_{\hat U X}=\{(u_{ki,x_j}): B_{k,ji}\ne 0, k=1,..., p\}$. Let $E_{u_{ki}}\subseteq E_{\hat UX}$ be the set of edges incident to $u_{ki}$ in ${\cal B}(\hat A', \hat B)$. By the relation between generic rank and the bipartite matching (\citep[Prop 2.1.12]{Murota_Book}), it is readily known that condition i) of Lemma \ref{switch-lemma} is satisfied, if and only if ${\cal B}(\hat A', \hat B)$ has a matching with size $n$. 

Similar to Section \ref{main-ILP}, introduce binary variables $t=\{t_{ki}: k=1,...,p, i=1,...,m_k\}$ and $y=\{y_{uv}: (u,v)\in E_{\hat XX}\cup E_{\hat UX}\}$, and we can formulate ${\cal P}_4$ and ${\cal P}_5$ as equivalent ILPs. 
\begin{proposition}
	Suppose $(A_{\sigma(\cdot)}, B_{\sigma(\cdot)})$ is structurally controllable. Problems ${\cal P}_4$ and ${\cal P}_5$ are equivalent to the following ILPs ${\cal P}_{4}^{\rm ILP}$ and ${\cal P}_{5}^{\rm ILP}$, respectively, in the sense that, if $(y^\star,t^\star)$ is an optimal solution to the corresponding ILP ${\cal P}_{j}^{\rm ILP}$ ($j=4,5$), then $S^{\star}=\{u_{ki}: t^\star_{ki}=1, 1\le k \le p, 1\le i \le m_k\}$ is the optimal solution to the corresponding ${\cal P}_j$. 
	\begin{align}
		\min_{y,t} \quad & \sum \nolimits_{k=1}^p \sum \nolimits_{i=1}^{m_k} c_{ki}t_{ki} \tag{${\cal P}^{\rm ILP}_4$}\label{ILP4} \\
		{\rm{s.t.}}\quad & \sum \nolimits_{u:(u,v)\in E_{\hat XX}\cup E_{\hat UX}} y_{uv} = 1, \forall v\in X \label{C1-switched} \\
		& \sum \nolimits_{v:(u,v)\in E_{\hat XX}\cup E_{\hat UX}} y_{uv} \le 1, \forall u \in \hat X \cup \hat U \label{C2-switched}\\
		& \sum \nolimits_{k=1}^p \sum \nolimits_{j=1}^{m_k} w^k_{ij}t_{kj}\ge 1, \forall i\in \hat {\cal I} \label{C3-switched} \\
		& t_{ki} \ge \sum \nolimits_{(u,v)\in E_{u_{ki}}} y_{uv}, \forall k=1,...,p, i =1,...,m_k \label{C3-add-switched} \\
		& y_{uv}\in \{0,1\}, \forall (u,v)\in E_{\hat XX}\cup E_{\hat UX} \label{C4-switched} \\
		& t_{ki}\in \{0,1\}, \forall k=1,...,p, i =1,...,m_k.\label{C5-switched}
	\end{align}	
	\begin{align}
		\min_{y,t} \quad & \sum \nolimits_{k=1}^p \sum \nolimits_{i=1}^{m_k} c_{ki}t_{ki} \tag{${\cal P}^{\rm ILP}_5$}\label{ILP5} \\
		{\rm{s.t.}}\quad & (\ref{C1-switched}), (\ref{C2-switched}), (\ref{C3-switched}), (\ref{C3-add-switched}),  (\ref{C4-switched}), {\rm and} \ (\ref{C5-switched}) \\
		& \sum \nolimits_{i=1}^p \sum \nolimits_{j=1}^{m_i} t_{ij}\le k \label{C2-switched-swit}
	\end{align}
\end{proposition}

\begin{proof}
	Similar to the proof of Proposition \ref{ILP-formulation}, constraints (\ref{C1-switched}), (\ref{C2-switched}) and (\ref{C4-switched}) ensure that there is a maximum matching that matches $X$ in ${\cal B}(\hat A', \hat B)$. This means condition i) of Lemma \ref{switch-lemma} is met. Constraints (\ref{C3-switched}) and (\ref{C5-switched}) ensure that each source-SCC of ${\cal G}(\hat A)$ is input-reachable. Furthermore,  constraint (\ref{C3-add-switched}) indicates that the input $u_{ki}$ is selected if any edge of $E_{u_{ki}}$ is contained in the maximum matching of ${\cal B}(\hat A', \hat B)$ associated with constraints (\ref{C1-switched}), (\ref{C2-switched}) and (\ref{C4-switched}). 
	Hence, both conditions of Lemma \ref{switch-lemma} are satisfied with the constraints (\ref{C1-switched})-(\ref{C5-switched}). Additionally, constraint (\ref{C2-switched-swit}) yields that the cardinality constraint is met. Optimizing the objective functions of ${\cal P}_{4}^{\rm ILP}$ and ${\cal P}_{5}^{\rm ILP}$ certainly leads to the optimal solutions to ${\cal P}_{4}$ and ${\cal P}_{5}$.
\end{proof}

%\begin{remark}
%	A variant of ${\cal P}_5$ is to determine the minimum cost of inputs to achieve structural controllability for the switched system (*) while the total number of inputs (for all the modes) does not exceed 
%	a prescribed number $k\in {\mathbb N}_+$. The ILP formulation of this problem could be obtained from ${\cal P}_5^{\rm ILP}$ by changing constraint (\ref{C2-switched-swit}) to  $\sum \nolimits_{i=1}^p \sum \nolimits_{j=1}^{m_i} t_{ij}\le k$. 
%\end{remark}

It is easy to see that, upon letting $k=\sum \nolimits_{i=1}^p m_i$, constraint (\ref{C2-switched-swit}) will become redundant and ${\cal P}^{\rm ILP}_5$ reduces to ${\cal P}^{\rm ILP}_4$, indicating ${\cal P}_4$ is a special case of ${\cal P}_5$. As such, in what follows, we will focus on ${\cal P}_5$, and the obtained results are directly applied to ${\cal P}_4$.

\begin{definition}[Joint SSSI constraint] For $(A_\sigma(\cdot), B_{\sigma(\cdot)})$ in (\ref{plant-switched}), it satisfies the joint SSSI constraint, if $(\hat A, \hat B)$ satisfies the SSSI constraint. 
\end{definition}

Notably, the joint SSSI constraint does {\emph{not}} require each subsystem $(A_i, B_i)$ ($i=1,...,p$) to satisfy the SSSI constraint (see Example \ref{example-switched}). This is because two vertices belonging to the same source-SCC in ${\cal G}(\hat A)$ may come from different source-SCCs in ${\cal G}(A_i)$. On the other hand, if each subsystem $(A_i, B_i)$ satisfies the SSSI constraint, $(\hat A, \hat B)$
will automatically satisfy the joint SSSI constraint, since the source-SCCs of ${\cal G}(\hat A)$ must consist of unions of source-SCCs of ${\cal G}(A_i)$. 

\begin{theorem} \label{theorem-switch}
	Suppose $(A_{\sigma(\cdot)}, B_{\sigma(\cdot)})$ is structurally controllable and satisfies the joint SSSI constraint. If ${\cal P}_5$ is feasible, then the following LP-relaxation ${\cal P}_{5}^{\rm LP}$ of ${\cal P}_{5}^{\rm ILP}$ always has an integral optimal solution corresponding to the optimal solution to ${\cal P}_5$. 
	\begin{align}
		\min_{y,t} \quad & \sum \nolimits_{k=1}^p \sum \nolimits_{i=1}^{m_k} c_{ki}t_{ki} \tag{${\cal P}^{\rm LP}_5$}\label{LP5} \\
		{\rm{s.t.}}\quad & (\ref{C1-switched}), (\ref{C2-switched}), (\ref{C3-switched}), (\ref{C3-add-switched}),  {\rm and} \ (\ref{C2-switched-swit}) \\
		& 0\le y_{uv}\le 1, \forall (u,v)\in E_{\hat XX}\cup E_{\hat UX}  \\
		& 0\le t_{ki} \le 1, \forall k=1,...,p, i =1,...,m_k.
	\end{align}
	Consequently, ${\cal P}_5$ (as well as ${\cal P}_4$) is polynomially solvable with the joint SSSI constraint.  
\end{theorem}

To prove Theorem \ref{theorem-switch}, as we have argued in Section \ref{main-result}, it suffices to demonstrate that the constraint matrix of ${\cal P}_{5}^{\rm LP}$ is TU. To this end, associated with
$(\hat A', \hat B)$, let matrix $\tilde M \in \{0,\pm 1\}^{((p+1)n+2\sum \nolimits_{i=1}^p m_i+ \hat r+ 1)\times (n_{\hat E}+ \sum \nolimits_{i=1}^p m_i)}$ be constructed in the same way as (\ref{M-hat-eq}), in which $E_{XX}\cup E_{UX}$ is replaced with $E_{\hat XX}\cup E_{\hat U X}$, $X_L\cup U \cup X_R$ with $X\cup \hat X \cup \hat U$, $w$ with $[w^1,..., w^p]$, and $\alpha$ with $[0_{1\times n_{\hat E}}, 1_{1\times \sum \nolimits_{i=1}^p m_i}]$, where $n_{\hat E}=|E_{\hat XX}\cup E_{\hat U X}|$. We have the following proposition, which yields Theorem \ref{theorem-switch}. 

%Additionally, let matrix $\tilde \alpha \in \{0,1\}^{p\times (n_{\hat E}+ \sum \nolimits_{i=1}^p m_i)}$ be
%$$\tilde \alpha=\left[\begin{array}{ccccc}
	%0_{1\times n_{\hat E}} & 1_{1\times m_1} & 0_{1\times m_2} & \cdots & 0_{1\times m_p} \\
	%0_{1\times n_{\hat E}} & 0_{1\times m_1} & 1_{1\times m_2} & \cdots & 0_{1\times m_p} \\
	%\vdots & \vdots & \vdots & \ddots & \vdots \\
	%0_{1\times n_{\hat E}}  & 0_{1\times m_1} & 0_{1\times m_2} & \cdots & 1_{1\times m_p}
	%\end{array}\right].    $$
	%After that, define matrix $\bar M$ as 
	%$$\bar M=  \left[\begin{array}{c}
		%\tilde M	\\
		%\tilde \alpha	
		%\end{array}\right]. $$

		\begin{proposition} \label{TU-switched-size}
			If $(A_\sigma(\cdot), B_{\sigma(\cdot)})$ satisfies the joint SSSI constraint, then the above constructed $\tilde M$ is TU. 
		\end{proposition}
		
		\begin{proof} Note that $[w^1,..., w^p]$ has the same structure as $w$ under the joint SSSI constraint (i.e., each column has at most one nonzero entry $1$). Hence, similar reasoning to the proof of Proposition \ref{TU-M} yields that $\tilde M$ is TU. %The details are omitted due to their similarities.    %Consequently, to show the TU of $\bar M$, it suffices to show that every submatrix of $\bar M$ that contains a submatrix of $\tilde \alpha$ is TU. To this end, consider
		\end{proof}
		
		Following the spirit of Proposition \ref{TU-switched-size} and Theorem \ref{extend-theorem}, we have the following corollary to generalize Theorem \ref{theorem-switch}.  
		\begin{corollary} \label{re-TU-switch}
			If $[w^1,..., w^p]$ satisfies the restricted TU condition, then both ${\cal P}_4$ and ${\cal P}_5$ can be solved in polynomial time via solving the respective LP relaxations of ${\cal P}_4^{\rm ILP}$ and ${\cal P}_5^{\rm ILP}$.
		\end{corollary}
		
		\begin{remark}
			Remarkably, the restricted TU of $[w^1,..., w^p]$ does not imply that, for each mode, the source-SCC-input incidence matrix associated with $(A_i,B_i)$ is restrictedly TU (although $w^i$ should be so for each $i$), nor the reverse, the latter of which is distinct from the joint SSSI constraint.   %different from the joint SSSI constraint 
		\end{remark}
		
		Finally, consider a switched system where the input structure is {\emph{fixed}} during the switching, i.e.,  $\tilde B_{\sigma(\cdot)}$ in (\ref{plant-switched}) is replaced with a time-invariant $\tilde B$. In other words,  $\tilde B_{1}=\cdots=\tilde B_p=\tilde B$, recalling $p$ is the number of switching modes. Such a scenario may occur, for example, in networked systems with switching topologies but fixed inputs \cite{hou2016structural}. Denote the associated structured system of  $(\tilde A_{\sigma(\cdot)}, \tilde B)$ by  $(A_{\sigma(\cdot)},B)$. Corresponding to ${\cal P}_4$ and ${\cal P}_5$, we consider two input selection problems for $(A_{\sigma(\cdot)},B)$, that is, selecting the minimum cost of inputs $B({\cal J})$, and selecting the minimum cost of inputs $B({\cal J})$ with a cardinality upper bound on $|{\cal J}|$, both to achieve structural controllability of $(A_{\sigma(\cdot)}, B({\cal J}))$, ${\cal J}\subseteq \{1,...,m\}$. Denote these problems as ${\cal P}^{\rm fix}_4$ and ${\cal P}^{\rm fix}_5$, respectively. The following corollary reveals, ${\cal P}^{\rm fix}_4$ and ${\cal P}^{\rm fix}_5$ can be solved in polynomial time provided the source-SCC-input incidence matrix of $(\hat A, B)$ is restrictedly TU, with $\hat A$ defined above.
		
		\begin{corollary}
			If the source-SCC-input incidence matrix of $(\hat A, B)$ is restrictedly TU, then ${\cal P}^{\rm fix}_4$ and ${\cal P}^{\rm fix}_5$ can be solved in polynomial time.
		\end{corollary}
		
		\begin{proof}
			According to \cite{Z.S2002Controllability}, $(\tilde A_{\sigma(\cdot)}, \tilde B_{\sigma(\cdot)})$ is controllable, if and only if the controllability matrix $C(\tilde A_{\sigma(\cdot)}, \tilde B_{\sigma(\cdot)})$ defined as follows has full row rank:
			$$\begin{array}{l}  C(\tilde A_{\sigma(\cdot)}, \tilde B_{\sigma(\cdot)})= [ \tilde B_1,..., \tilde B_p, \tilde A_1\tilde B_1,...,\tilde A_p\tilde B_1,...,\tilde A_p\tilde B_p,\\
				\tilde A_1^2\tilde B_1,..., \tilde A_p\tilde A_1\tilde B_1,...,\tilde A_1^2\tilde B_p,...,\tilde A_p\tilde A_1\tilde B_p,..,\tilde A_1^{n-1}\tilde B_1,...,\\
				\tilde A_p\tilde A_1^{n-2}\tilde B_1,...,\tilde A_1\tilde A_p^{n-2}\tilde B_p,...,\tilde A_p^{n-1}\tilde B_p].
			\end{array} $$
			Under the condition that $\tilde B_1=\cdots= \tilde B_p=\tilde B$, upon defining $\{\tilde B'_{\sigma(\cdot)}\}$ as $\tilde B'_{1}=\tilde B, \tilde B'_2=\cdots = \tilde B'_{p}=0_{n\times m}$, it is easy to see that ${\rm rank} C(\tilde A_{\sigma(\cdot)}, \tilde B_{\sigma(\cdot)})= {\rm rank} C(\tilde A_{\sigma(\cdot)}, \tilde B'_{\sigma(\cdot)})$. Hence, ${\cal P}^{\rm fix}_4$ and ${\cal P}^{\rm fix}_5$ can reduce to ${\cal P}_4$ and ${\cal P}_5$ with the newly defined $\{\tilde B'_{\sigma(\cdot)}\}$. The proposed statement then follows directly from Corollary \ref{re-TU-switch}. 
		\end{proof}
		
		%can reduce to ${\cal P}_4$ and ${\cal P}_5$, respectively. Consequently, they 
		%\section{Comparisons with other methods}	
		
		\section{Illustrative examples }		 \label{example-sec}

		\begin{figure}
			\centering
			\includegraphics[width=1.9in]{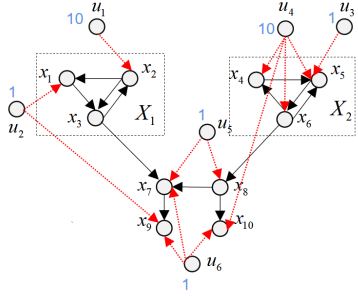}\\
			\caption{System digraph of $(A,B)$ in Section \ref{example-sec}. Dotted red edges represent the input links, with the numbers in blue near each input vertex being its cost.}\label{example}
		\end{figure}
		
		\begin{figure}
			\centering
			\includegraphics[width=3.3in]{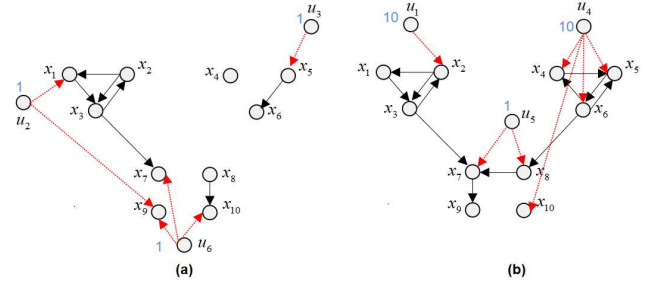}\\
			\caption{System digraphs ${\cal G}(A_1,B_1)$ (a) and ${\cal G}(A_2,B_2)$ (b) of the two subsystems of the switched system in Example \ref{example-switched}. }\label{switch-example}
		\end{figure}
		
		We provide two examples here to illustrate the effectiveness of the proposed methods. 
		
		\begin{example} \label{example-fixed}
			Consider system $(A,B)$ with its system digraph ${\cal G}(A,B)$ given in Fig. \ref{example}, which has $6$ inputs, $10$ states, and $26$ edges. The cost of inputs is $c=[c_1,...,c_6]=[10,1,1,10,1,1]$. This system contains two source-SCCs, with $X_1=\{x_1,x_2,x_3\}$ and $X_2=\{x_4,x_5,x_6\}$. Hence, $r=2, m=6$. As no inputs connect with $X_1$ and $X_2$ simultaneously, the SSSI constraint is met. The corresponding matrix $$w=\left[\begin{array}{cccccc}
				1 &  1 & 0 & 0 & 0 & 0 \\
				0 & 0 & 1 & 1 & 0 & 0
			\end{array}\right].$$ Construct LPs ${\cal P}_1^{\rm LP}$ and ${\cal P}_2^{\rm LP}$ with decision variables $(y,t)$, $y\in {\mathbb R}^{26}$ and $t\in {\mathbb R}^{6}$.  					
			
			Using the Matlab LP solver {\emph{linprog}} to solve the LP ${\cal P}_1^{\rm LP}$ associated with this system, we obtain $t^\star=[0,1,1,0,1,1]$, implying the optimum $\sum\nolimits_{i=1}^6 c_it^\star_i=4$. This means the optimal solution is $S^{\star}=\{u_2,u_3,u_5,u_6\}$, and the corresponding minimum cost is $4$. Remarkably, this result is consistent with the one obtained via the LP-rounding.
			
			Next, we solve the LP ${\cal P}_2^{\rm LP}$ with $k=3$ to obtain the integral optimal solution (see footnote \ref{foot2} on how to obtain the integral solution from a fractional one). And we obtain $t^\star=[0,1,0,1,1,0]$, with the optimum $\sum\nolimits_{i=1}^6 c_it^\star_i=12$. This means the optimal solution with a cardinality upper bound $3$ is $S^{\star}=\{u_2,u_4,u_5\}$, and the corresponding cost is $12$.
			Finally, we set $c=[1,1,1,1,1,1]$ and then ${\cal P}_1$ reduces to determining the minimum number of inputs to achieve structural controllability. We obtain $t^\star_i=1$ for $i=1,4$ and $t^\star_i=0$ otherwise. Hence, the minimum number of inputs
			for structural controllability is $2$, with the optimal solution $S^{\star}=\{u_1, u_4\}$. If we consider the original input cost $c=[10,1,1,10,1,1]$, this solution has cost $20$, which can also be obtained by setting $k=2$ in ${\cal P}_2$.  \hfill $\square$
		\end{example}
		
		\begin{example} \label{example-switched}
			Consider a switched system $(A_{\sigma(\cdot)},B_{\sigma(\cdot)})$ with two switching modes. The system digraphs of subsystems $(A_1, B_1)$ and $(A_2,B_2)$ are given in Figs. \ref{switch-example}(a) and (b), which implies $m_1=m_2=3$. Let the cost of inputs be $c=[c_1,...,c_6]=[10,1,1,10,1,1]$ (i.e., the same as in Example \ref{example-fixed}).   On the one hand, for subsystem $(A_1, B_1)$, $u_2$ connects with the source-SCCs $\{x_1,x_2,x_3\}$ and $\{x_9\}$, while for subsystem $(A_2, B_2)$, $u_4$ connects with the source-SCCs 
			$\{x_4,x_5,x_6\}$ and $\{x_{10}\}$ simultaneously. This indicates each subsystem does not satisfy the SSSI constraint. On the other hand,  it is easy to check that the system digraph of the corresponding $(\hat A, \hat B)$ is exactly Fig. \ref{example}. Therefore, the whole system $(A_{\sigma(\cdot)},B_{\sigma(\cdot)})$ satisfies the joint SSSI constraint. By Theorem \ref{theorem-switch},  we could adopt the LP-relaxations to solve ${\cal P}_4$ and ${\cal P}_5$ associated with this system. 
			
			To this end, let us build ${\cal P}_5^{\rm LP}$ with $k=3$. Solving ${\cal P}_5^{\rm LP}$ yields $t^{\star}=[0,1,1,0,1,0]$ with the corresponding cost $\sum \nolimits_{i=1}^6 c_it^{\star}_i=3$. Hence, this minimum cost switched 
			input selection with no more than $3$ inputs is that, selecting $\{u_2,u_3\}$ for subsystem $(A_1, B_1)$ and $\{u_5\}$ for subsystem $(A_2, B_2)$. If we set $k=2$, then solving ${\cal P}_5^{\rm LP}$ returns the solution $t^{\star}=[0,1,0,1,0,0]$, with the cost $\sum \nolimits_{i=1}^6 c_it^{\star}_i=11$. This implies, selecting $\{u_2\}$ for subsystem $(A_1, B_1)$ and $\{u_4\}$ for subsystem $(A_2, B_2)$ incurs the minimum cost 
			$11$ with no more than $2$ inputs.  \hfill $\square$
		\end{example}
		
		Comparing Examples \ref{example-fixed} and \ref{example-switched}, we find: 1) both for a fixed (non-switched) and a switched system, with a bigger cardinality upper bound, the cost of the obtained solution tends to be smaller. This highlights the significance of the cardinality-constrained minimum cost input selections ${\cal P}_2$ and ${\cal P}_5$; 2) even though the `union' of subsystems (i.e., $(\hat A, \hat B)$) of the switched system is the same as the fixed system, the former tends to have smaller input costs for achieving structural controllability under the same cardinality constraint. This is consistent with the fact that switched systems are often more efficient (in terms of the number of inputs, control energy, etc.) to be controlled than the non-switched ones \cite{li2017fundamental}.

	%	\section{Conclusions} \label{conclusion}
			\section{Conclusions} \label{conclusion}
		In this paper, we explore polynomially solvable conditions for the (cardinality-constrained) minimum cost input selection problems 
		both for non-switched and switched structured systems. Though the NP-hardness in general, we reveal that if the input structure satisfies certain regulations, characterized by the restricted TU of the input-source-SCC incidence matrix, irrespective of the connections between each input and the states within the same source-SCC or from the non-source-SCCs, those problems are polynomially solvable via solving the corresponding LP-relaxations. A particular case is the SSSI constraint, which often emerges in some practical systems, and has been extended to the switched systems, resulting in the joint SSSI constraint. In the general case, we obtain some lower and upper bounds for the considered problems via LP-relaxation and LP-rounding. It is still unclear how to solve those problems using graph-theoretic algorithms under the addressed conditions, which could be the future work, perhaps with the help of the LP primal-dual algorithms \cite{lawler2001combinatorial}. %In the future, apart from the LP-relaxation, it is interesting to develop graph-theoretic algorithms for those problems under the addressed conditions, perhaps with the help of the LP primal-dual algorithms \cite{lawler2001combinatorial}.
		
		% while preserving the polynomial solvability % does not require each subsystem to satisfy the SSSI constraint
		%	This paper investigates two related constrained input selection problems for structural controllability. Instead of giving approximation algorithms, we provide polynomially solvable conditions for them.
		%	We first formulate these problems as equivalent ILPs. We then reveal that under the said SSSI constraint, those ILPs could be solved efficiently by their LP relaxations using the off-the-shelf LP solvers. This is achieved by proving that the corresponding constraint matrices of the ILPs are TU. Our condition contains almost all the existing known nontrivial polynomially solvable ones as special cases. In the future, it is interesting to develop graph-theoretic algorithms for the addressed problems, perhaps with the help of the primal-dual algorithms for the corresponding LPs \cite{lawler2001combinatorial}.

		%How to drop Assumption \ref{broader-assumption} and explore the complexity status of ${\cal P}_i|_{i=1}^3$ in the general case could be the future work.

		\section*{Appendix: Proof of restricted TU}
	%	\subsection{Proof of restricted TU}
		
		{\bf{Proof of the extended SSSI constraint:}} Under this constraint, $w$ is a block-diagonal matrix. It suffices to show that every nonzero diagonal block of $w$, which is a matrix with all entries being $1$, denoted as $1_{r'\times m'}$, is restrictedly TU. Since every square submatrix of $1_{(r'+1)\times m'}$ has a determinant being either $0$ (if the dimension is greater than $1$) or $1$ (if the dimension is $1$), $1_{r'\times m'}$ is restrictedly TU by definition. %This finishes the proof.
		
		{\bf{Proof of the row-monotone and column-monotone constraints:}} First, we show for any $p\in {\mathcal N}_+$, the following matrix is restrictedly TU:
		$${\cal W}_p=\left[\begin{array}{cccc}
			1 & 1 & \cdots & 1 \\
			0 & 1 & \cdots & 1 \\
			\vdots & \vdots & \ddots & \vdots \\
			0  & 0 & \cdots & 1
		\end{array}\right]\in \{0,1\}^{p\times p}. $$
		Since ${\cal W}_p$ contains a row $1_{1\times p}$, it suffices to show ${\cal W}_p$ is TU. Consider any square submatrix ${\cal W}'$ of ${\cal W}_p$. As ${\cal W}_p$ is row-monotone and column-monotone, so is ${\cal W}'$.   Hence, ${\cal W}'$ either contains a zero column or two identical columns, or is upper triangular with all diagonal entries being $1$. This means, $\det {\cal W}'\in \{0,1\}$, proving that ${\cal W}_p$ is TU. Consequently, any sub-matrix of ${\cal W}_p$ is restrictedly TU by definition. Now consider a row-monotone $w\in \{0,1\}^{r\times m}$. If all rows of $w$ are non-decreasing, then each row must be a row of ${\cal W}_m$.  Remove all the repeated rows from $w$ and do some row permutations on the resulted matrix, and we can obtain a matrix $w'$ that is a sub-matrix of ${\cal W}_m$. Note as proved, $w'$ is restrictedly TU. Since row (as well as column permutations) will not change the absolute value of determinants, and removing repeated rows will not affect the property of being restrictedly TU (by definition), it turns out that $w$ is restrictedly TU. The case that all rows of $w$ are non-increasing follows a similar way, and so does the case that $w$ is column-monotone.

		{\bf{Proof of the permutable row/column-monotone constraint:}} Since row and column permutations will not change the absolute value of determinants, the restricted TU follows directly from the fact that row (column)-monotone $w$ is restrictedly TU.

			\section*{\refname}
		\bibliographystyle{elsarticle-num}
		{\small
			\bibliography{yuanz3}
		}
	\end{document}